\documentclass[usenatbib,usegraphicx]{mn2e}

\usepackage{amssymb}
\usepackage{ulem}
\newcommand{\aap}{A\&A}
\newcommand{\aaps}{A\&AS}
\newcommand{\aapr}{A\&AR}
\newcommand{\aj}{AJ}
\newcommand{\apj}{ApJ}

\newcommand{\apjs}{ApJS}
\newcommand{\mnras}{MNRAS}
\newcommand{\pasp}{PASP}

\newcommand{\araa}{ARA\&Ap}

\defcitealias{C99}{C99}
\defcitealias{FS90}{FS90}

\begin{document}

\title[Low-luminosity galaxies in the NGC~5044 Group]{Evolutionary properties 
of the low-luminosity galaxy population in the NGC~5044 Group\thanks{Based on 
observations collected at the European Southern Observatory, La Silla (Chile)
and at the {\sc Casleo} Observatory, operated under agreement between CONICET, UNLP, UNC, and UNSJ,
Argentina}}

\author[Buzzoni et al.]{A. Buzzoni$^{1}$, S. A. Cellone$^{2,3}$,
P. Saracco$^{4}$ \& E. Zucca$^{1}$\\
$^1$INAF - Osservatorio Astronomico di Bologna, Via Ranzani 1, 40127 Bologna,
Italy\\
{\sf alberto.buzzoni@oabo.inaf.it, elena.zucca@oabo.inaf.it}\\
$^2$Facultad de Ciencias Astron\'omicas y Geof\'\i sicas, Universidad Nacional
de La Plata, Paseo del Bosque, B1900FWA La Plata, Argentina\\
$^3$IALP, CCT La Plata, CONICET-UNLP\\
{\sf scellone@fcaglp.unlp.edu.ar}\\
$^4$INAF - Osservatorio Astronomico di Brera, Via Brera 28, 20121 Milano,
Italy\\
{\sf paolo.saracco@brera.inaf.it}\\
 }

%\offprints{}

\date{Received ; Accepted}

%\authorrunning{S.\ A.\ Cellone \& A.\ Buzzoni}
%\titlerunning{Low luminosity galaxies in the NGC 5044 Group}
\maketitle

\label{firstpage}

\begin{abstract}
With this third paper of a series we present Johnson-Gunn $B,g,V,r,i,z$ multicolour 
photometry for 79 objects, including a significant fraction of the faintest galaxies 
around NGC~5044, assessing group membership on the basis of apparent morphology (through 
accurate S\'ersic profile fitting) and low-resolution ($R = 500-1000$) optical 
spectroscopy to estimate redshift for 21 objects.

Early- and late-type systems are found to clearly separate in the S\'ersic 
parameter space,  with the well-known luminosity vs. shape relation being mostly
traced by different morphological types spanning different ranges in the
shape parameter $n$. 
A significantly blue colour is confirmed for Magellanic irregulars 
(Sm/Im's), while a drift toward bluer integrated colours is also an issue for 
dE's. Both features point to a moderate but pervasive star-formation activity even 
among nominally ``quiescent'' stellar systems.
Together, dE's and Im's provide the bulk of the galaxy luminosity function, 
around $M(g) \simeq -18.0\pm 1.5$, while the S0 and dSph components dominate, 
respectively, the bright and faint-end tails of the distribution. This special mix
places the NGC~5044 group just ``midway'' between the high-density cosmic 
aggregation scale typical of galaxy clusters, and the low-density environment of 
looser galaxy clumps like our Local Group. 
The bright mass of the 136 member galaxies with available photometry and morphological
classification, as inferred from appropriate M/L model fitting, 
amounts to a total of $2.3\,10^{12}$~M$_\odot$. 
This is one seventh of the total dynamical mass of the group, according to its X-ray emission. 
Current SFR within the group turns to be about 23~M$_\odot$~yr$^{-1}$, 
a figure that may however be slightly increased facing the evident activity 
among dwarf ellipticals, as shown by enhanced H$\beta$ emission in their spectra.

Lick narrow-band indices have been computed for 17 galaxies, probing all the 
relevant atomic and molecular features in the 4300-5800~\AA\ wavelength range.
Dwarf ellipticals are found to share a sub-solar metallicity ($-1.0 \lesssim [\mathrm{Fe/H}] 
\lesssim -0.5$) with a clear decoupling between Iron and $\alpha$ elements, as
already established also for high-mass systems. Both dE's and dS0's are consistent 
with an old age, about one Hubble time, although a possible bias, toward higher 
values of age,  may be induced by the gas emission affecting the H$\beta$ strength.
\end{abstract}

%%%%%%%%%%%%%%%%%%%%%%%%%%%%%%%%%%%%%%%%%%%%%%%%%%%%%%%%%%%%%%%%%%%%%%%%%%

\section{Introduction \label{s_i}}

Low-mass and low-surface brightness (LSB) galaxies, including in this definition 
both dwarf ellipticals and irregulars, have since long surged to a  
central role in the general debate about galaxy formation and evolution. 
These objects, typically characterized by an absolute blue magnitude fainter than 
$M_B \simeq -16$ and a mean surface brightness 
$\langle \mu_\mathrm{e}\rangle \gtrsim 22$~mag~arcsec$^{-2}$ \citep{fb94}, display in fact 
a quite wide range of morphological and spectrophotometric properties 
\citep[e.g.][]{jerjen00} such as to rise the question whether they assume an 
individual role as a sort of pristine building blocks to form more massive systems 
\citep[as in the CDM theoretical scheme, e.g.][]{kaufmann93}, or they rather 
originate as coarse ``by-products'' of the formation of ``standard'' ellipticals 
and spirals \citep{knezek99,lisker07}.

The problem of the genetic signature of low-mass systems is not a negligible one as, 
although in a more silent way, dwarf galaxies display in many cases continuous star 
formation activity, which makes them possible efficient ``engines'' to enrich the 
interstellar medium \citep[e.g.][]{arimoto95}.
Actually, a major problem in this framework resides in the apparent dichotomy between 
active ongoing star formation, widely observed in LSB galaxies, and the exceedingly 
low metallicity of their present-day stellar populations \citep{arimoto90}. The most 
striking case in this context is surely that of IZw~18, as recently discussed by 
\citet{aloisi07}. Again, while for some dwarf ellipticals we might be dealing 
with a somewhat quiescent evolutionary scenario, reminiscent of old globular cluster 
history, in other situations a ``tuned'' star formation could have proceeded allover 
the entire galaxy life \citep[e.g.][]{grebel05,held05} still leaving, nowadays, an 
important fraction of fresh (primordial?) gas \citep[e.g.][]{carrasco95,kuzio04}.

As far as galaxy spatial distribution is concerned, the ubiquitous presence of LSB irregulars 
in the different cosmic environments \citep{cowie91,fb94,saracco99} is clearly at odds with the 
morphology-density relation \citep{dressler80}, which has been proven to hold for 
``standard'' galaxies for a wide range of densities and scales 
\citep{postman84,maia90,hp03,kelm05}. 
A systematic observational work to map the LSB galaxy distribution in selected 
zones of the sky has been carried out by different teams, leading to complete 
surveys and morphological catalogues of some loose groups of galaxies and nearby 
clusters \citep{binggeli85,karachentseva85,ichikawa86,davies88,ferguson89,fs90,jd97,sh97}.

These morphological surveys serve as a basic reference to any further 
spectrophotometric analysis of individual galaxies 
\citep{bm88,bh91,cellone94,held94,secker97,cellone99,smith08}.
However, efficient detection and accurate observation of LSB galaxies still remain a 
formidable task, even for HST observations \citep[e.g.][]{ferrarese06} especially when 
moving to distances beyond the Virgo and Fornax Clusters, the two best studied aggregates 
so far \citep[e.g.][]{hilker99,drinkwater01,conselice01,deady02,mieske04,gavazzi05,sabatini05}.

A complementary view may be gained through the study of smaller groups. In fact, 
the wide range of densities which characterizes these aggregates makes them the 
natural link between galaxy clusters and the field. Detailed mapping of the distribution 
and nature of low-mass galaxies in small groups will fill the ``void'' between fully relaxed systems 
\citep[like actually the case of NGC 5044 group, e.g.][]{forbes07} to very loose and 
possibly unbound aggregates \citep{campos04,roberts04}. Besides probing the 
properties of dwarf galaxies in different environments, these have the advantage 
that even a moderate-sized sample would be fairly representative of the whole group
population. At the same time, depth effects would be minimized allowing a
better analysis of distance-dependent quantities.

In this work we want to further follow up a long-range study on the low-mass LSB 
galaxy population of the NGC~5044 group. In particular, after assessing group 
morphology and membership in previous contributions \citep{cb01,cb05}, we would 
like to tackle here the problem of galaxy evolutionary properties, according to a 
multiwavelength observational approach that includes integrated and surface-brightness 
multicolour photometry, further complemented in most cases with a spectroscopic
input. Our analysis will rely on original stellar population synthesis models, 
to which refer galaxy data in order to lead to a convenient and physically 
self-consistent picture of the galaxy evolutionary status for the NGC~5044 group. 

In this framework, we will arrange our discussion by presenting
the observed database in Sec.~2. Its morphological characterization will be carried out 
in Sec.~3 by fitting galaxy surface brightness with a \citet{S68} profile. These results in most 
cases refine our previous discussion in \citet{cb05} leading also to a homogeneous 
morphological classification of the full galaxy population in the NGC~5044 group. 
Together with up-to-date redshift measurements, the revised morphological parameters proved
to be a useful additional piece of information to constrain group membership 
and eventually assess the overall physical properties of the galaxy group as a whole.
As a central step in this direction, in Sec.~4 we study galaxy distribution in different 
magnitude and colour domains.
Starting from apparent magnitudes and morphological types, a match with the \citet{buzzoni05}
template galaxy models also allows us to lead to a realistic estimate of galaxy stellar 
mass and a reference birthrate figure for the galaxy group as a whole.

The spectroscopic properties for our sample are finally discussed in some detail in Sec.~5, 
where Lick narrow-band indices are derived for most of the observed galaxies. The diagnostic
diagrams compare $\alpha$-element features, like Magnesium and Calcium, versus Iron and 
Balmer line stregth, also including the contribution from Carbon molecules like CH (G band)
and  C$_2$. We will discuss our results and summarize our main conclusions in Sec.~6.

\section{Observed database}

Our galaxy sample consists of a total of 79 mostly dwarf and LSB galaxies, mainly 
selected from the \citet{fs90} catalogs, and representative of the faint 
galaxy population surrounding the standard elliptical NGC~5044, in the range 
$-18 \la M_B \la -11$ mag,\footnote{For the NGC~5044 group a distance modulus 
$(m-M) = 31.96-5\log(h_{100})$ can be estimated, according to the 
mean spectroscopic redshift of all group members in \citet{cb05}.
For $H_0 = 75$~km\,s$^{-1}$ Mpc$^{-1}$ this leads to a value of 
$(m-M) = 32.58$~mag, that we will adopt throughout this paper.} 
regardless of morphology. Johnson $BV$ photometry and $griz$ imaging in the 
Gunn system \citep{tg76,wheh79,sgh83}, as well as mid-resolution ($R = 500$-1000) 
spectroscopy for a subsample of 24 objects have been collected during 
different observing runs from 1996 to 2000. Specifically, 40 objects have been 
observed during two observing campaigns (on the nights of Apr 16-17, 1999 and Apr 29-May 1, 2000) 
at the {\sc Eso} 3.6m telescope of La Silla (Chile), while previous extended observations at the 2.15m
telescope of the {\sc {\sc Casleo}} observatory, in San Juan (Argentina) were carried out along three 
observing runs (on the nights of May 10-13 1996, Apr 8-10 1997, and Mar 19-21 1999) 
providing supplementary data for 57 galaxies, 18 of which in common with the 
{\sc Eso} sample.\footnote{Here, and throughout the paper, the prefix ``N'' in our name 
coding of galaxies stands for the catalogue number in \citet{fs90}. A supplementary 
letter ``A'', ``B'', or ``C'' to the standard catalog number refers to the newly discovered LSB 
galaxies of \citet{cellone99,cb05}, and this paper, maintaining the reference 
number of the closest bright galaxy.\label{footn}}

\subsection{Imaging}

ESO observations have been carried out with EFOSC2, equipped with a Loral 2k CCD 
in a $2\times 2$ binned mode providing a platescale of 0.32~arcsec pixel$^{-1}$. 
The $5.3\arcmin \times 5.3\arcmin$ field of view allowed in most cases to image 
more than one galaxy within each frame. A total of 24 fields 
were imaged in the $griz$ bands \citep[see Fig.~1 in][for the exact location map]{cb05}, 
including 33 known low-luminosity galaxies, plus the bright SbI-II NGC\,5054, 
NGC~5044 itself, and 6 newly discovered LSB galaxies \citep[see][]{cb05}.
Atmospheric conditions were photometric all the way, with sub-arcsec seeing during
the first run, and a poorer performance (FWHM $\ga 1.5$~arcsec) in the 2000 run.
In case of multiple exposures, a final image of the field was obtained
by stacking the individual frames after all processing steps were completed. 
Standard stars from the lists of \citet{sgh83} and \citet{jorgensen94} were also observed 
during each run for calibration purposes.

Johnson $BV$ imaging at the {\sc Casleo} telescope has been carried out, on
the other hand, during two runs in 1996 and 1999. The telescope was equipped with a Tektronix 
1024$\times$1024 CCD, providing a 9$\arcmin$ circular field of view in direct-imaging mode, 
with a platescale of 0.83~arcsec pixel$^{-1}$.
Standard stars for magnitude calibration were selected from the list of \citet{landolt92}.
Seeing conditions were typically poorer than {\sc Eso} observations, ranging between 2.0
and 2.8 arcsec. Partial results from the 1996 data have been presented in \citet{cellone99}.

Image processing of both {\sc Eso} and {\sc Casleo} data was done using \textsc{iraf}\footnote{IRAF 
is distributed by the National Optical Astronomy Observatories, which are operated by the
Association of Universities for Research in Astronomy, Inc., under cooperative
agreement with the National Science Foundation.}, complemented with a few of
our own \textsc{fortran} routines. Each bias-corrected frame was twilight flat-fielded,
while residual fringe patterns on {\sc Eso} $i$ and $z$ images were corrected, mostly
within a final $\sim 1$\% accuracy level, by subtraction 
of the appropriate templates \citep[see][for more details on the procedure]{cb05}. 
Cosmic rays were cleaned up using the \textsc{iraf} task \textsc{cosmicrays}, and
the sky background, fitted with a tilted plane, was finally subtracted to obtain the
final images of each field.

The photometric reduction procedure yielded a nominal figure ($\sigma \sim 0.001$~mag) for 
the internal magnitude uncertainty of bright galaxies, being the total photometric error 
dominated throughout by the external uncertainty in the zero-point calibration. Overall, for a 
$g \sim V \sim 16$ galaxy we estimate a full magnitude uncertainty of the order of
$\sigma \sim 0.03$~mag, a figure that may raise to $\sigma \sim 0.06$~mag
for the faintest ($g \sim V \sim 21$) objects ($\sigma \sim 0.10$~mag for {\sc Casleo}
data).

A check of internal self-consistency of the $BV$ photometry has been made possible 
by the combined observations of galaxies N49 ($V = 16.5$) and N83A ($V = 20.2$),
in common to the 1996 and 1999 runs. For the first bright object, 
the coincidence was excellent, as both $B$ and $V$ isophotal magnitudes and surface 
brightness have been reproduced within $\sim 0.004$~mag, and $\sim 0.01$~mag~arcsec$^{-2}$, 
respectively. Much poorer is the match of N83A, instead, a very faint dSph galaxy. 
In the latter case, in fact, repeated $V$ photometry resulted in a nearly negligible offset
of 0.06~mag, that raised however to 0.35~mag for the $B$ band. A similar figure has been 
found for the surface-brightness measurements, with a difference of 0.002 and 
0.45~mag~arcsec$^{-2}$ in $V$ and $B$, respectively. This warns of the fact that 
photometric parameters of very faint galaxies may actually show considerable errors.

\subsubsection{Johnson to Gunn magnitude conversion}

In order to ease a more homogenous discussion of the whole galaxy sample 
one would like to convert {\sc Casleo} $BV$ photometry into {\sc Eso} Gunn system. 
To this aim we took advantage of the 18 galaxies in common between the two 
subsamples, as summarized in Table~\ref{t1}. For these galaxies,
integrated magnitudes within one $g$-effective radius, as identified by the 
S\'ersic fitting profile (see Table~\ref{a2}), have been computed matching 
{\sc Casleo} $BV$ and {\sc Eso} $gr$ photometry.

\begin{figure}
\centerline{
\includegraphics[width=0.9\hsize,clip=]{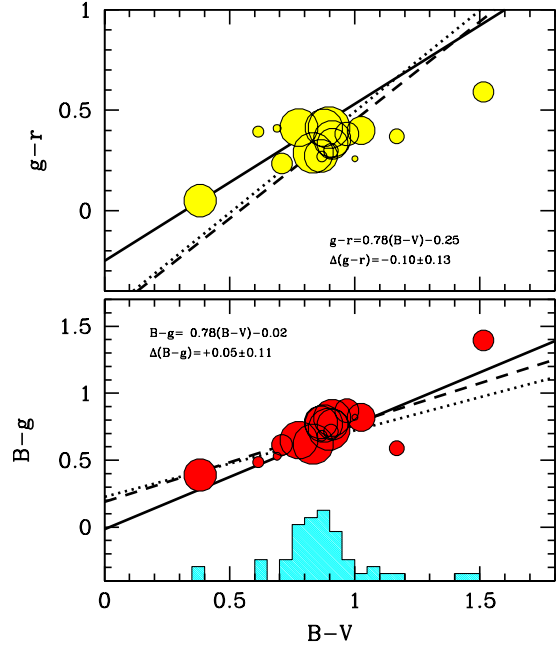}
}
\caption{
Gunn vs.\ Johnson system transformation for the 18 galaxies in common between
ESO and {\sc Casleo} samples. For each object, $g, r, V$, and $B$ aperture magnitudes
(and the corresponding colours) refer to a fixed effective radius $\rho_\mathrm{e}^{[S]}$ as from 
the $g$-band S\'ersic fitting profile, according to Table~\ref{t1}.
Dotted and dashed lines are the relations of transformation by \citet{kent85}
and \citet{jorgensen94}, respectively, while solid line is the adopted 
\citet{buzzoni05} calibration. The latter is explicitely reported in each panel, 
together with the mean uncertainty in converting $(B-V)$ to $(g-r)$ and $(B-g)$ colours.
Marker size in both panels is proportional to galaxy V luminosity. Note the
bunched distribution of the calibrating galaxies, that closely traces the
$(B-V)$ distribution of the whole {\sc Casleo} sample, as sketched by the histogram in the 
lower panel.}
\label{f1}
\end{figure}

The linear fit to these data is reported in Table~\ref{t2}. One has to 
remark, however, that the strong clustering of the data in the colour 
domain (see Fig.~\ref{f1}), poorly constrains the relations of 
transformation. Alternatively, more standard sets of empirical conversion 
relationships can be found in the literature, mainly relying on the 
observation of stellar grids. This is the case, for instance, of the works 
by \citet{jorgensen94} and \citet{kent85}, also reported in Table~\ref{t2}.
A further set of fully theoretical conversion equations has been 
considered in the table, relying on the \citet{buzzoni05} template galaxy 
models. The latter calibration may in  principle be more suitable for our 
aims, as it better accounts for the slightly different locus of stars and galaxies 
in the colour domain, given the ``broader'' spectral energy distribution 
(SED) of the latter.

\begin{table}
%\centering
\scriptsize
\caption{Galaxy sample in common to {\sc Casleo} and {\sc Eso} observing runs}
\label{t1}
\begin{tabular}{lrcccc}
\hline
Name & \multicolumn{1}{c}{$\rho_\mathrm{e}^S$}  & \multicolumn{4}{c}{Aperture magnitude} \\
        &  \multicolumn{1}{c}{[arcsec]}  &   g   &     r   &  V   &   B   \\
\hline
N30  & 10.90 &  15.939 &  15.526 &  15.835 &  16.712 \\
N34  &  7.19 &  16.829 &  16.496 &  16.684 &  17.595 \\
N42  & 17.49 &  15.723 &  15.435 &  15.508 &  16.342 \\
N49  &  8.11 &  16.472 &  16.422 &  16.479 &  16.861 \\
N50  &  7.16 &  15.938 &  15.524 &  15.811 &  16.588 \\
N54  & 18.36 &  16.487 &  16.214 &  16.404 &  17.269 \\
N54A &  4.32 &  21.569 &  21.310 &  21.388 &  22.389 \\
N55  &  5.26 &  20.151 &  19.757 &  20.021 &  20.636 \\
N56  &  7.07 &  19.481 &  19.182 &  19.291 &  20.196 \\
N62  &  7.76 &  19.840 &  19.470 &  19.261 &  20.429 \\
N70  & 10.76 &  17.810 &  17.428 &  17.711 &  18.680 \\
N70A &  5.45 &  20.997 &  20.587 &  20.837 &  21.527 \\
N75  &  7.74 &  16.009 &  15.652 &  15.915 &  16.825 \\
N83  & 12.52 &  17.300 &  16.901 &  17.094 &  18.120 \\
N83A &  7.20 &  20.356 &  20.086 &  20.170 &  21.039 \\
N109 &  7.30 &  18.301 &  18.066 &  18.207 &  18.916 \\
N153 & 12.04 &  15.475 &  15.061 &  15.308 &  16.206 \\
B3   &  2.41 &  18.346 &  17.755 &  18.226 &  19.741 \\
\hline
\end{tabular}
\end{table}

\begin{table}
%\centering
\caption{$BV$ vs.\ $gr$ conversion relationships}
\label{t2}
\begin{tabular}{ll}
\hline
\hline
\multicolumn{2}{l}{{\bf Least-square fit to the 18 galaxies in common}} \\
g--r = 0.33\,(B--V) +0.05 & \quad ($\sigma,\rho$) = (0.08 mag,~0.68) \\
\phantom{~~~~~~~~~} $\pm 9$\phantom{~~~~~~~~~~~~\,} $\pm 8$ & \\
B--g = 0.78\,(B--V) +0.03  & \quad ($\sigma,\rho$) = (0.10 mag,~0.86) \\ 
\phantom{~~~~~~~~\,} $\pm 11$\phantom{~~~~~~~~~~~} $\pm 10$ & \\
\hline
\multicolumn{2}{l}{{\bf Buzzoni (2005)}} \\
g--r = 0.78\,(B--V) --0.25  &  ~~($\Delta,\sigma$) = (--0.10 mag, 0.13 mag) \\
B--g = 0.78\,(B--V) --0.02 &  ~~($\Delta,\sigma$) = (0.05 mag, 0.11 mag)\\
\hline
\multicolumn{2}{l}{{\bf J\o rgensen (1994)}} \\
g--r = 1.01\,(B--V) --0.52   & ~~($\Delta,\sigma$) = (--0.03 mag, 0.18 mag) \\
B--g = 0.50\,(B--V) +0.23 & ~~($\Delta,\sigma$) = (0.06 mag, 0.13 mag)\\
\hline
\multicolumn{2}{l}{{\bf Kent (1985)}} \\
g--r = 0.98\,(B--V) --0.53    & ~~($\Delta,\sigma$) = (0.00 mag, 0.17 mag)\\
B--g = 0.59\,(B--V) +0.19  & ~~($\Delta,\sigma$) = (0.01 mag, 0.12 mag)\\
\hline
\end{tabular}
\end{table}

The output of the different transformation sets is compared with our galaxy 
distribution in Fig.~\ref{f1}. None of the literature relations seems to 
suitably account for galaxy conversion along the entire colour range. 
On the other hand, one has also to consider that the bunching group 
of common objects in Table~\ref{t1} is fairly representative
of the bulk of {\sc Casleo} galaxies, as well (see the overplotted histogram in the 
lower panel of Fig.~\ref{f1}), and the few deviating targets at the most 
extreme colour edges are in fact among the faintest objects in the sample, 
possibly with highest internal uncertainty. 
To a closer analysis, and within the large scatter of the observations,
one can notice a marginally more consistent match with the \citet{buzzoni05}
calibration, and we eventually decided to 
adopt this equation set for our {\sc Casleo} sample to be converted to Gunn 
magnitudes. According to Fig.~\ref{f1}, the (unweighted) 
external uncertainty provided by the \citet{buzzoni05} conversion equations 
(in the sense ``observed'' -- ``predicted'') amounts to 
$\Delta (g-r) = -0.10\pm 0.13$ and $\Delta (B-g) = +0.05 \pm 0.11$~mag. 

\subsection{Spectroscopy}

EFOSC2 has also been used in spectroscopic mode for {\sc Eso} mid-resolution 
observations for a sub-sample of 24 galaxies (including NGC~5044 itself, 
the central galaxy of the group). Spectra along the 
4300-6400~\AA\ wavelength interval were obtained through a 1 arcsec longslit with {\sc Eso} 
grism \#8 and a 6~\AA\ FWHM resolution ($R \simeq 1000$) sampled at a scale 
of 2.1~\AA\,pixel$^{-1}$. When possible, a second target galaxy was included on the 
same spectrum frame of the main object. When no catalogued galaxy could be 
used as second target, we considered any likely background galaxy within 
the slit; six such objects were observed in this way \citep[see][for details]{cb05}.

The {\sc Eso} sample includes virtually all (three out of four) galaxies observed 
at {\sc Casleo} during the 1997 run in spectroscopic mode. The {\sc Casleo} 
observations were carried out with  a Boller \& Chivens spectrograph equipped 
with a 300~line mm$^{-1}$ dispersor grid, providing a FWHM resolution of 9.5~\AA\ 
($R \simeq 500$) along a bluer wavelength range (namely, from 3500--5700~\AA) 
compared to {\sc Eso}, sampled at bins of 4.4~\AA\,pixel$^{-1}$.

Each 2-D spectrum in our galaxy sample was bias-subtracted and flat-fielded 
via calibration spectra of an halogen lamp.
One-dimensional spectra were extracted using \textsc{iraf}  standard routines, 
and wavelength-calibrated by means of He-Ne-Ar lamp technical exposures. 
Flux calibration of {\sc Casleo} and {\sc Eso} (1999 run) spectra were carried out by means
of the observation of spectrophotometric standard stars from the \citet{baldwin84} 
catalog. Calibration of {\sc Eso} 2000 observations relied instead on standard stars 
from the list of \citet{gutierrez88}.

Finally, individual 1-D spectra were coadded to obtain the fiducial spectrum 
of each target. Cross-correlation with the spectrum of NGC~5044, used as a 
template, provided eventually a measure of the redshift of each object in the 
ESO sample, as we already discussed in \citet{cb05}. A redshift of 
$cz = 2351 \pm 71$~km\,s$^{-1}$ was derived, independently, for galaxy N29, 
the only one observed at {\sc Casleo} with no {\sc Eso} counterpart.

\section{Morphological parameters}

An accurate morphological study for 33 galaxies in our sample has already been 
undertaken in \citet{cb05}, providing a (re)classification and membership
estimate of most objects on the basis of their image properties and, wherever 
possible, of the available spectroscopic redshift. Taking previous analysis as 
a reference, we want here to complete our review for the whole galaxy set. 

We took advantage for our classification of the ``objective'' shape parameters, as provided
by a \citet{S68} fit of radial surface brightness profile ($\mu(\rho)$), i.e.
\begin{equation}
\mu = \mu_0 +1.086\left({\rho\over\rho_0}\right)^n.
\label{eq:sersic}
\end{equation}
In this scheme, three parameters constrain galaxy morphology, namely a 
central surface brightness ($\mu_0$), a scale radius ($\rho_0$), and a ``shape index''
$n$ (being, as a reference, $n=0.25$ for the standard de Vaucouleurs profile),
which discriminates between ``spiky'' ($n<1$) and ``cored'' ($n>1$) galaxies.
While many bright dE/dS0 do show composite structure 
\citep[i.e.\ ``bulge''+``disk'', see][]{cellone99,cb01}, we did not attempt here to
make any profile decomposition; S\'ersic shape parameter $n$ thus traces the 
global morphology of each galaxy.
All the calculations made use of the $g$-band or the $V$-band 
observations for the {\sc Eso} and {\sc Casleo} subsamples, respectively.

Given the different PSFs between the {\sc Eso} and {\sc Casleo}
observations, instead of trying to correct for seeing effects, we chose a
simpler ---and safer--- approach, following \citet{cellone99}, who showed
that seeing induced errors on the fitted Sersic parameters tend to be
negligible when an inner cutoff radius $\approx 2 \times$\,FWHM is used
for the fits \citep[see also][]{gavazzi05}. An outer cutoff at a radius
where $S/N \simeq 1$ prevents against systematic errors due to an
imperfect sky subtraction.
While the choice for the inner cutoff turned out to have no impact on 
the {\sc Eso} data, thanks to their superior seeing, some 
limitation might be expected, in principle, for the fit of {\sc Casleo} 
galaxies, one third of which being comprised with their scale radius
within $2 \times$\,FWHM. 

A summary of the results for the whole galaxy sample is displayed in Tables~\ref{a2} and 
\ref{a3}.\footnote{Note that among the {\sc Casleo} galaxies, a more careful inspection of 
the observed fields also led us to discover 6 more LSB galaxies (all consistent with a 
dSph morphology), not reported ever. For reader's better convenience, their coordinates 
are summarized in Table~\ref{a1} of the Appendix.}

In addition to the fitting morphological parameters ($\mu_0$, $\rho_0$ and $n$ appear, 
respectively in col.~8, 9, and 10 of the tables), we also provided a more direct estimate 
of galaxy size, assumed to coincide with the isophotal radius ($\rho_{27}$) 
at $\mu(g) = 27$~mag\,arcsec$^{-2}$, and of effective radius ($\rho_\mathrm{e}$),
which encircles 50\% of the luminosity within $\rho_{27}$ (see cols.\ 6 and 7 in 
the tables, respectively).\footnote{The 
aproximation of $\rho_{\rm gal}$ with $\rho_{27}$ might lead of course to a 
slightly underestimated galaxy total luminosity, an effect mostly contained 
within $\sim 0.2$~mag, as shown by \citet[][see Fig.\ 5 therein]{cb05}.}

\begin{figure}
\centerline{
\includegraphics[width=0.85\hsize,clip=]{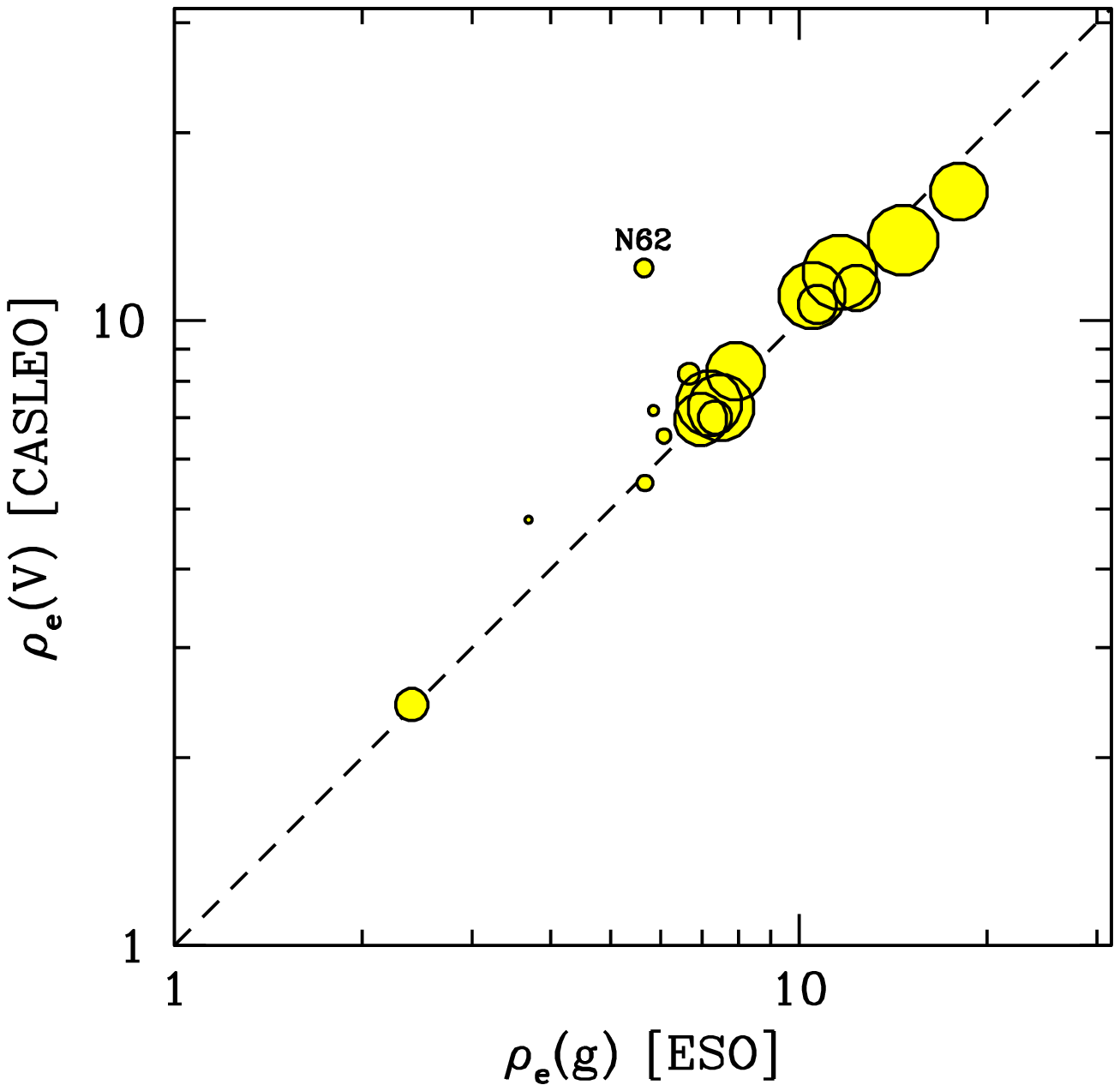}
}
\centerline{
\includegraphics[width=0.85\hsize,clip=]{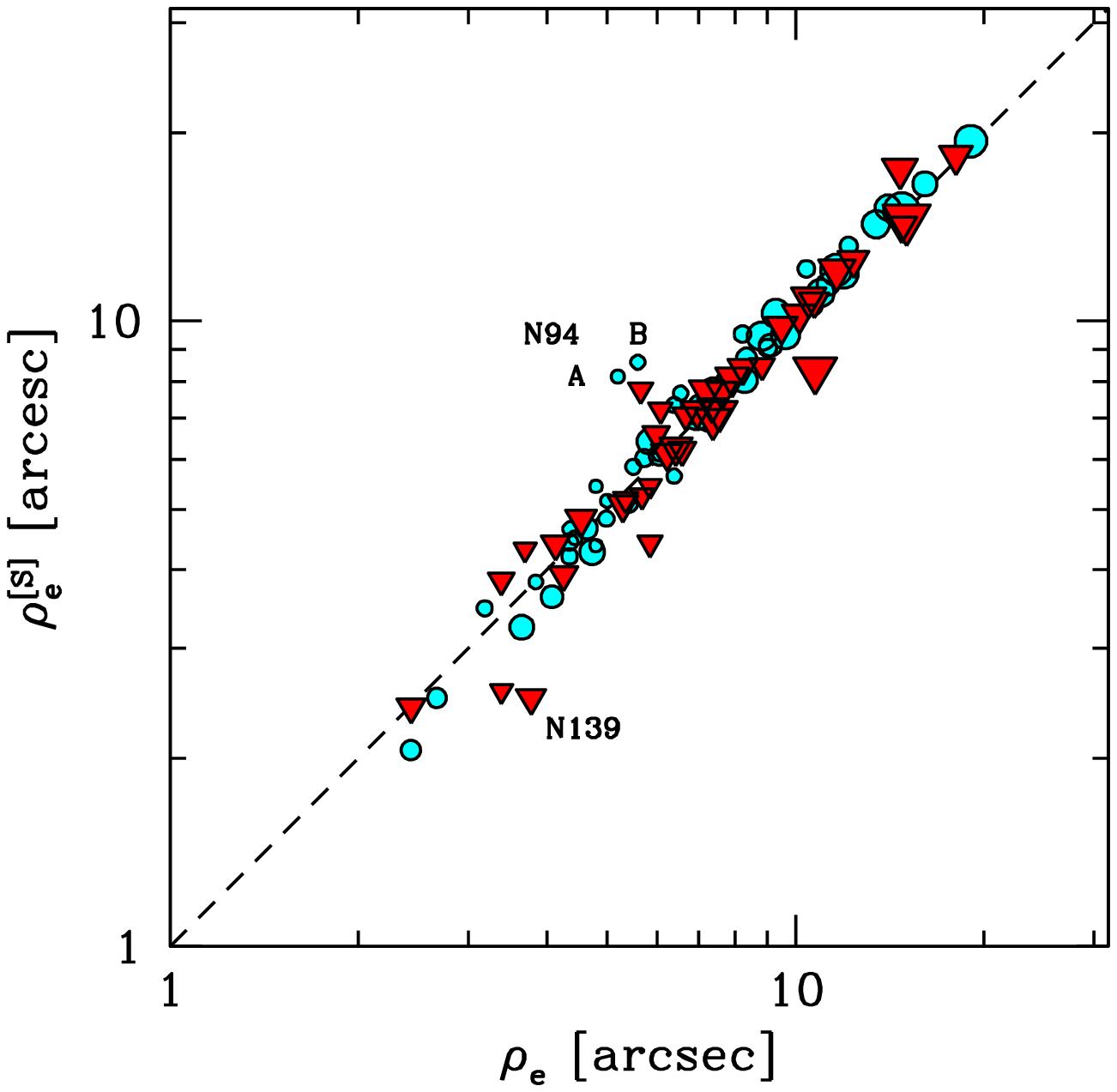}
}
\caption{
{\it Upper panel:} A combined comparison of the observed effective radius $\rho_\mathrm{e}$
(in arcsec) in the $V$ and $g$ bands for the 18 galaxies in common between {\sc Eso} and 
{\sc Casleo} samples.
The two measures are consistent within a $\pm 11$~\% relative uncertainty.
{\it Lower panel:} Observed ($\rho_\mathrm{e}$) vs.\ computed ($\rho_\mathrm{e}^{[S]}$) values of the 
effective radius, the latter as from the S\'ersic fitting profile, for the whole 
(ESO+{\sc Casleo}) sample of 79 galaxies. The two estimates agree within a $\pm 9$~\% 
relative scatter.
Marker size is proportional to galaxy $g$ or $V$ luminosity, respectively,
with {\sc Casleo} objects plotted as dots, and {\sc Eso} objects as triangles. 
See the text for a brief discussion of the few outliers in both panels.
}
\label{f2}
\end{figure}

The 18 galaxies in common between the {\sc Eso} and {\sc Casleo} subsamples provided a useful cross-check, 
allowing us to consistently compare the $V$ vs.\ $g$ fits. For the measured 
values of $\rho_\mathrm{e}$, this is shown in the upper panel of Fig.~\ref{f2}, where 
one can appreciate a nice relationship in place between the two sets of 
observations with just one evident outlier, that is galaxy N62 among the 
faintest and most vanishing objects in the sample \citep[see Fig.~2 in][]{cb05}.
After rejecting this object, we have that {\sc Casleo} effective radii match
the corresponding {\sc Eso} figures within a $\pm 11$\% relative uncertainty.
An external check of the accuracy in the fitting procedure can be obtained from
the analysis of the lower panel of Fig.~\ref{f2}, where the ``observed'' value of $\rho_\mathrm{e}$
is matched with the corresponding parameter inferred from the S\'ersic fit 
($\rho_\mathrm{e}^{[S]}$).\footnote{Given a S\'ersic scale length parameter, $\rho_0$, the
fitting effective radius $\rho_\mathrm{e}$ can be defined as $\rho_\mathrm{e}/\rho_0 = w^n$, where
the parameter $w = w(n)$ comes from the (numerical) solution of the relevant relation 
$\Gamma(2/n)/2 = \gamma(2/n,w)$, with $\Gamma$ and $\gamma$ being respectively the complete 
and incomplete Gamma functions.}
The two estimates agree within a $\pm 9$\% relative scatter, after a 
3-$\sigma$ rejection of three relevant outliers, namely the two dSph galaxies
N94A and B (from the {\sc Casleo} sample) and the (background?) elliptical
N139 (from the {\sc Eso} sample), as singled out in the figure.
Regarding the shape parameter $n$, which is particularly sensible to differing seeing 
conditions and sky-subtraction errors, we also obtain a good agreement 
($\langle \Delta n \rangle = -0.07 \pm 0.04$) between {\sc Eso}-{\sc Casleo} samples when the 12 brighter 
($g_{27}< 18$) galaxies are considered,  while a poorer match 
($\langle \Delta n \rangle = 0.23 \pm 0.12$) is attained for the 6 fainter systems.

It is of special interest to explore the $n$-index distribution across the
Hubble/de Vaucouleurs morphological type of our galaxies. The latter has 
been inferred from a careful ``eye-driven'' standard classification 
of the original images, complementing the previous \citet{cb05} results
(see the ``CB05'' column in Tables~\ref{a2} and \ref{a3}). Given a better observing material, our 
types refine and update, in most cases, the original scheme by \citet{fs90} 
(column ``FS90'' in the tables).
At the same time, our revised morphological classification led membership re-assessment 
for galaxies with no redshift available \cite[see][]{cellone07}.
A new set of radial velocity measurements, recently provided by \citet{mendel08},
raised to 45 the number of galaxies with available redshift in our sample. In most cases,
the morphologically assigned membership status (see the corresponding code, $m_c$, in col.~2
of Tables~\ref{a2} and \ref{a3}) was confirmed by the new redshifts,
while just for three galaxies it had to be changed (from $m_c = 3$ to $m_c = 1$),
leading also to changes in their respective morphological types (as from col.~5 of 
Tables~\ref{a2} and \ref{a3}).

\begin{figure}
\centerline{
\includegraphics[width=\hsize,clip=]{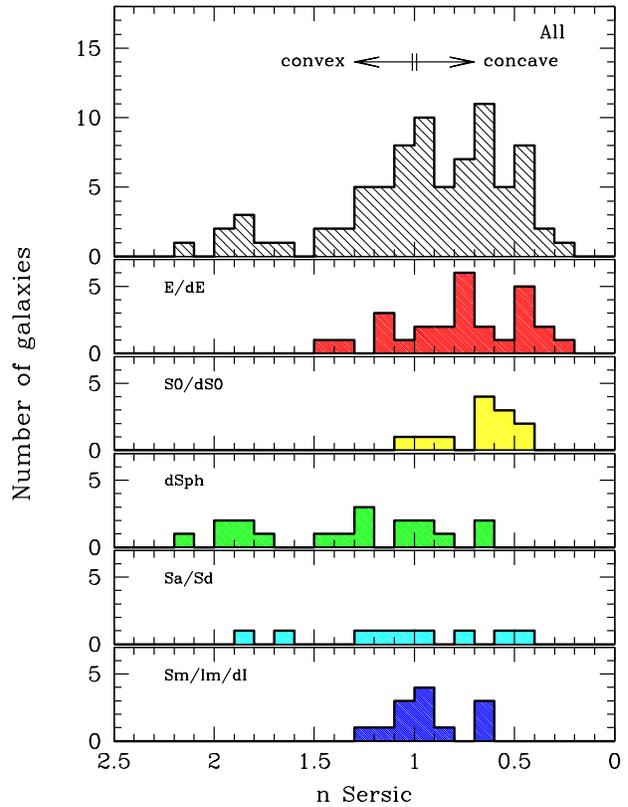}
}
\caption{
Galaxy distribution according to the S\'ersic shape fitting parameter ``$n$'' of 
eq.~(\ref{eq:sersic}), as reported in Table~\ref{a2} and \ref{a3}. For the
galaxies with both {\sc Casleo} and {\sc Eso} observations the latter data have been used.
Compared to a perfect exponential case ($n = 1$), a ``convex'' surface-brightness radial 
profile has to be expected for $n>1$, while a ``concave'' one is generated for 
$n<1$, as labelled in the top panel, that collects the whole sample of 79 objects. 
The lower panels disaggregate the galaxy population according to the 
morphological classification. See text for a discussion.
}
\label{f3}
\end{figure}

\begin{figure}
\centerline{
\includegraphics[width=\hsize,clip=]{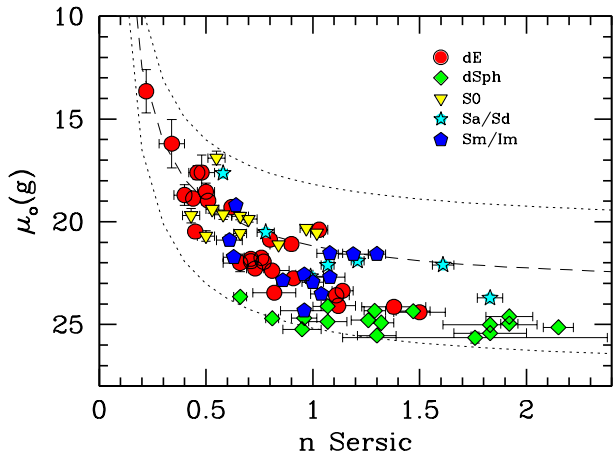}
}
\caption{
The central surface brightness ($\mu_0(g)$) is compared with the shape parameter $n$,
both as from the $g$- or $V$-band S\'ersic fitting profile for the whole galaxy
sample (see Table~\ref{a2} and \ref{a3}). 
In lack of {\sc Eso} observations, {\sc Casleo} $V$ data have been used, after converting to 
$g$ magnitudes according to the adopted calibration of Table~\ref{t2}.
The three overplotted curves mark the locus for fixed mean effective surface brightness,
namely $\langle \mu g_\mathrm{e}\rangle = 23$ mag\,arcsec$^{-2}$ (middle dashed curve), and
$\langle \mu g_\mathrm{e} \rangle = 20$ and 27 mag\,arcsec$^{-2}$ (upper and lower dotted curves,
respectively). Galaxy marker depends on morphological type, as labelled on the plot.
Note a clear morphological segregation of the different groups, with a prevailing
presence of dSph's among the LSB galaxy component, while ellipticals 
are among the brightest members approaching the standard de Vaucouleurs profile for 
$n \to 0.25$.
}
\label{f4}
\end{figure}

The match of the refined morphological types with the S\'ersic shape index is 
summarized in Fig.~\ref{f3}. As 
expected, the $n$-index properly accounts for the disk vs.\ bulge relative 
contribution to galaxy luminosity. In particular, bulge-dominated systems 
(E/dE+S0/dS0) are shown to display, on average, a ``concave'' 
($n<1$) surface-brightness profile, while disk-dominated systems (Sa/Sd)
span the full range of the $n$ index according to a wider disk/bulge 
luminosity partition. A shallower brightness profile also characterizes dwarf 
spheroidal (dSph) systems, where a cored structure likely leads to 
a ``convex'' ($n>1$) shape index. A pure exponential profile ($n \simeq 1$) is, 
on the contrary, the distinctive feature of later-type galaxies (Im/dI), that lack 
any clear nuclear condensation. To much extent, such a morphological 
segregation for the S\'ersic shape index is the ultimate responsible for the 
observed trend of $n$ with galaxy luminosity \citep[e.g.][]{cellone99}. 
As we will see in the next section, in fact, early-type (low-$n$) systems 
are among the brightest group members, while dSph (high-$n$) galaxies 
preferentially populate the faint-end tail of the group luminosity function. 
On a similar argument, the same effect neatly stems, as well, from
a study of the $\mu_0$ vs.\ $n$ S\'ersic parameters, like in
Fig.~\ref{f4}.

The plot shows, in addition, that early-type profiles actually 
tend to smoothly match the standard de Vaucouleurs case ($n \to 0.25$) as 
far as galaxy luminosity (and accordingly $\mu_0$) brightens up reaching 
the distinctive range of ``standard'' ellipticals, as confirmed by the 
straight $L-n$ observed relationship. At the faint end, dSph's display a wide 
range in $n$, while their central surface brightnesses remain almost constant 
at $\mu_0(g) \approx 25$~mag\,arcsec$^{-2}$.

\section{Photometric properties}

The Gunn $griz$ magnitudes and the Johnson $BV$ photometry provide a minimal
but effective set of measures to probe galaxy SEDs along the 4400-9000~\AA\ wavelength range. 
In addition to the morphological piece of information, in fact, a multicolour 
study of integrated luminosity of our targets, as well as of their surface brightness 
distributions could provide us with important clues to 
tackle the distinctive evolutionary properties of the galaxy population in 
the NGC~5044 group.

Again, our final results are collected in Table~\ref{a2} and \ref{a3},
respectively for the {\sc Eso} and {\sc Casleo} observations.\footnote{In spite of the 
higher internal uncertainty (see Sec.~2), note that magnitudes and colours in 
Tables~\ref{a2} and \ref{a3} are given with a nominal 3-digit precision simply 
for graphical purposes.} 
In both tables, the total apparent magnitude (encircled within the 
$\mu = 27$ mag\,arcsec$^{-2}$ isophote in the $g$ and $V$ bands, respectively) 
is reported in col.~11, together with the mean surface brightness within
the same isophotal radius, and within one effective
radius (cols.~12 and 13, respectively) according to the corresponding 
values of $\rho_{27}$ and $\rho_\mathrm{e}$.
Our output is finally completed with the $griz$ colours (cols.~14 to 19 in 
Table~\ref{a2}) and $(B-V)$ (cols.~14 and 15 in Table~\ref{a3}) across the 
same relevant apertures.
Unless explicitely stated, note that no correction for Galactic reddening 
has been introduced. According to \citet{burstein82}, the colour excess in 
the sky region around NGC~5044 amounts to $E(g-r) \simeq E(B-V) \sim 0.03$~mag, a figure
that may raise to $\sim 0.07$~mag according to the \citet{schlegel98}
reddening map.

\begin{figure}
\centerline{
\includegraphics[width=\hsize,clip=]{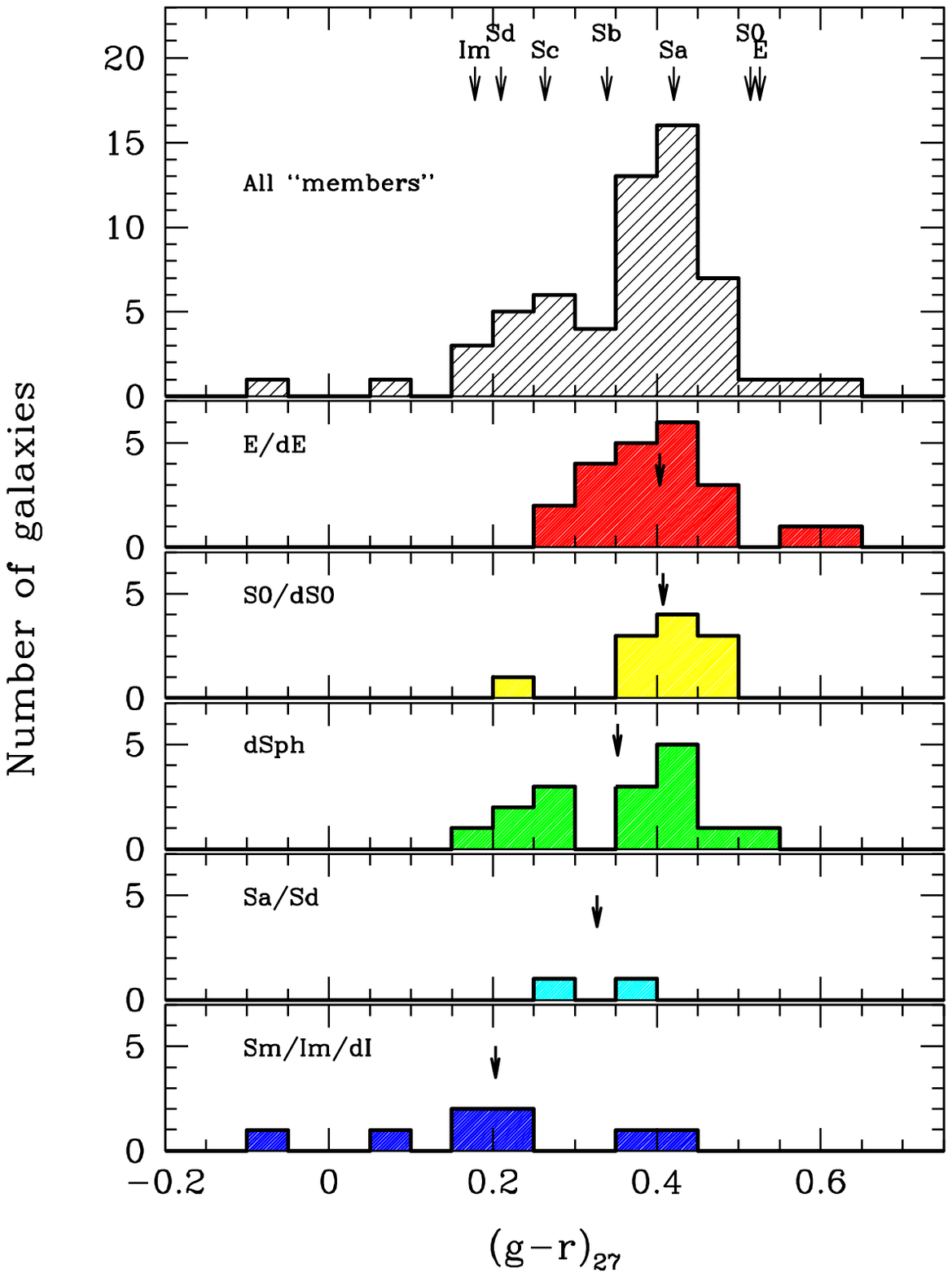}
}
\caption{
Apparent $(g-r)$ colour distribution of {\sc Eso}+{\sc Casleo} galaxy sample. 
Only likely members (``$m_c$'' code 1 and 2 of Table~\ref{a2} and \ref{a3}) are considered.
These 59 galaxies (upper panel) are disaggregated in the different morphological groups
along the different panels, as labelled. Mean colours as from the \citet{buzzoni05}
template galaxy models, matched to the observations according to \citet{schlegel98}
reddening estimate, are marked on the top by small arrows. These are to be compared with the mean observed 
colours of each morphological class (as from Table~\ref{t3}), 
as marked (small arrows in each panel).
}
\label{f5}
\end{figure}

\begin{table}
%\centering
\caption{Mean observed colours for the {\sc Eso}+{\sc Casleo} galaxy sample$^{(a)}$}
\label{t3}
%\scriptsize
\begin{tabular}{llr}
\hline\hline
\multicolumn{3}{c}{\hrulefill~{\sc Eso}+{\sc Casleo} sample~\hrulefill} \\
Type & $\langle(g-r)\rangle^{(b)}$ & no. of \\
   &  & galaxies \\
\hline
 dE     & 0.401{\tiny $\pm 0.082$} & 22 \\
 S0     & 0.408{\tiny $\pm 0.076$} & 11 \\
 dSph   & 0.352{\tiny $\pm 0.098$} & 16 \\
 Sa/Sd  & 0.327{\tiny $\pm 0.067$} & ~2 \\
 Sm/Im  & 0.203{\tiny $\pm 0.167$} & ~8 \\
        & \hrulefill   & \hrulefill \\
 All    & 0.360{\tiny $\pm 0.118$} & 59 \\ 
\hline
\hline
\noalign{$^{(a)}$ Only likely member galaxies are considered (i.e.\ with membership code 
$mc \le 2$ in Table~\ref{a2} and \ref{a3}). No correction for Galactic reddening 
has been applied.}
\noalign{$^{(b)}$ Mean colour within the $\mu(g)\equiv 27$~mag\,arcsec$^{-2}$ isophote.}
\end{tabular}
\end{table}

\subsection{Colour and magnitude distribution}

The colour distribution in the plane of integrated $(g-r)$ for the whole sample 
of 59 ``likely member'' (membership code $m_c \le 2$ in Tables~\ref{a2} and {\ref{a3}}) 
galaxies with available colours is displayed in the upper panel of Fig.~\ref{f5}.
In the other panels of the figure we also disaggregated the galaxy distribution
according to homogenous morphology groups.
Again, a clean overall trend can be recognized, with a colour shift from ``red''
to ``blue'' systems along the morphological sequence 
dE\,$\to$\,dS0\,$\to$\,dSph\,$\to$\,Sabc\,$\to$\,dI. The mean colours, according to
Table~\ref{t3}, are also marked in each panel, and consistently compare
with the theoretical predictions from \citet{buzzoni05} (the arrow sequence
in the upper panel of the figure) once considering the reddening shift.

Whithin the whole colour distribution of Fig.~\ref{f5}, one has to notice the 
key location of dwarf spheroidals, largely overlapping the apparent colour range of 
spirals and ellipticals but with slightly bluer colours compared to S0 systems. 
Like for the Local Group counterparts 
\citep[see, e.g.][for an updated review]{tht09}, this feature actually accounts
for the ``bivalent'' nature of these systems, bridging the morphological look 
of bulge-dominated systems and the composite photometric properties of 
disk-dominated galaxies, certainly reminiscent of a complex star 
formation history.

\begin{figure}
\centerline{
\includegraphics[width=\hsize,clip=]{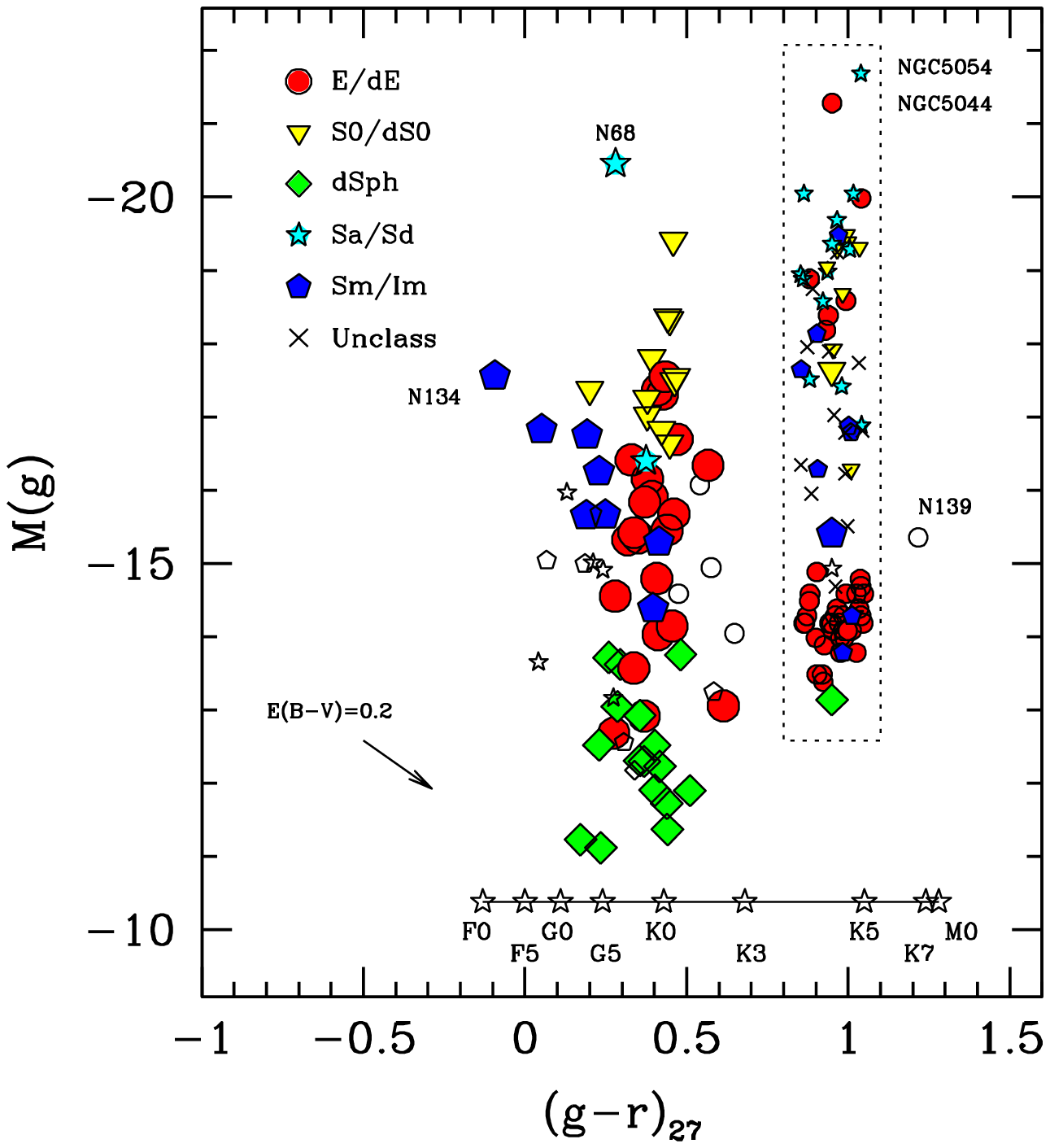}
}
\caption{colour-magnitude diagram for the 74 galaxies of the {\sc Eso}+{\sc Casleo} sample
with available $(g-r)$ or $(B-V)$ colour.
Solid markers refer to the likely group members (membership code $m_c \le 2$ 
in Table~\ref{a2} and \ref{a3}).
A supplementary set of five {\sc Casleo} galaxies from Table~\ref{a3}, plus 50 likely member 
galaxies from the \citet{fs90} original catalog, and 46 supplementary members from 
the \citet{mendel08} survey, all lacking the colour information,  
have been added within the dotted box (with little random scatter for better reading)
by converting $V$ or $B$ to $g$ magnitudes via Table~\ref{t2} calibration (see, in
addition, footnote \ref{foot9}).
A distance modulus $(m-M) = 32.58$~mag has been adopted to obtain absolute $g$
magnitudes. No correction for Galactic reddening has been applied to the data;
the reddening vector is however reported bottom left in the plot. Photometric
errors on $(g-r)$ colours  are between $\sigma(g-r) \sim 0.04$ and 0.06~mag  along 
the spanned magnitude range.
Some outstanding objects, discussed in the text, are singled out and labelled. 
Star markers in the bottom line indicate the reference colours for stars of 
different spectral type, according to \citet{straizys72}, as labelled.
}
\label{f6}
\end{figure}

The galaxy distribution across the $g$ vs.\ $(g-r)$ c-m diagram is displayed 
in Fig.~\ref{f6}. The full sample (79 galaxies) has been plotted, marking however
with solid symbols only the likely member galaxies (membership code $m_c \le 2$
in Tables~\ref{a2} and \ref{a3}).
In order to reach a better sampling of the overall luminosity function of the
cluster, we also picked up from the \citet{fs90} (50 galaxies) and \citet{mendel08} 
(46 galaxies) catalogs all the other likely or confirmed member galaxies (see
Table~\ref{t6} for their identification) not included in our observations. For 
these galaxies we converted the published $B$ photometry into $g$ magnitudes, 
according to our Table~\ref{t2} calibration. As no direct colour information is available 
for this subsample, we framed these galaxies in Fig.~\ref{f6} together with the 
five {\sc Casleo} objects lacking $(B-V)$ information.\footnote{For magnitude transformation 
of {\sc Casleo}, \citet{fs90} and \citet{mendel08} galaxies we adopted a reference 
$(B-V) \sim 0.8 \pm 0.3$. This reflects into a $g$ magnitude uncertainty of roughly 
0.3~mag in the framed points of Fig.~\ref{f6}.\label{foot9}}

Once considering the extended sample of all the \citet{fs90} and \citet{mendel08} galaxies,
the diagram shows an evident morphology/luminosity segregation within the different 
galaxy groups. As for the two leading systems (namely NGC~5044 itself, of type E, 
and the Sb spiral NGC~5054), the bright tail of the galaxy population actually 
consists of a balanced mix of standard spirals and ellipticals, while a prevailing
S0 population appears at absolute $g$ magnitudes around $M(g) \sim -18\pm 1.5$.
Dwarf ellipticals follow at fainter magnitudes, sharing the bulk of the galaxy 
population at intermediate luminosity ($M(g) \sim -15 \pm 2$) with dwarf irregulars. 
Definitely, the low-luminosity tail of galaxy distribution coincides with
a bunch of dwarf spheroidals, by definition, all fainter than $M(g) \gtrsim -14$ and also 
standing out for their extremely low surface brightness 
($\mu_0(g) \gtrsim 25$~mag~arcsec$^{-2}$, see Fig.~\ref{f4}). 
The dSph's clearly extend the colour-magnitude relation traced by early-type members 
down to $M(g) \approx -11$, with a scatter that is just mildly larger (likely due 
to their larger photometric errors) than the scatter at brighter magnitudes 
\citep[see][for an in-depth discussion on this topic]{smith11}.

Just on the basis of their location in the c-m diagram of Fig.~\ref{f6}, it is interesting
to note that most of the ``possible'' ($m_c = 3$ entries in Table~\ref{a2} and \ref{a3}) 
group members might in fact genuinely belong to the system.
If confirmed by spectroscopy, then the dE/Im/dSph component of the cluster 
could even further increase. However, this can certainly not be the case for 
galaxy N139, a clear outlier in the plot and more likely a background 
standard elliptical placed at $z \simeq 0.35$ according to its expected
k-corrected colours \citep[e.g.][]{buzzoni95}.

\begin{figure}
\centerline{
\includegraphics[width=\hsize,clip=]{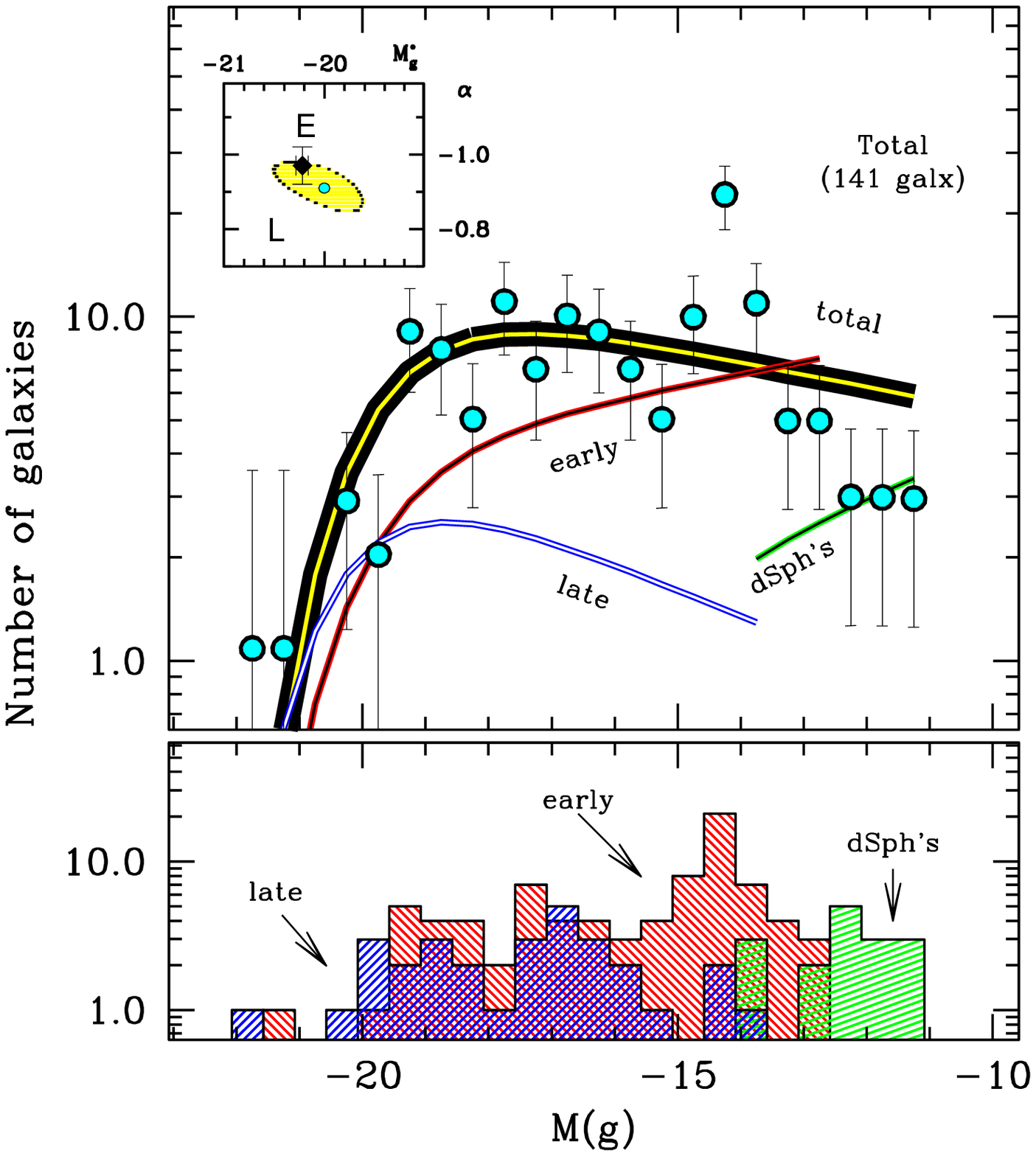}
}
\caption{
{\it Upper panel - } The $g$-band luminosity function for the 141 {\it bona fide} 
member galaxies (namely 59 objects from the present study, as from Table~\ref{t3}
plus 50 $m_c \le 2$ members from \citet{fs90} plus 32 spectroscopically confirmed members from 
\citet{mendel08} with available $B$ photometry. 
The thick solid line is the Schechter function for the global sample, according
to the STY method of \citet{sandage79}. An arbitrary normalization factor has been
applied to the curve such as to rescale to the observed 141 galaxies.
The global $\alpha - M^*_g$ fitting parameters, together with their $1\sigma$ 
confidence ellipse are also reported in the top left inset, comparing with
the \citet{zandivarez10} general figures from the SDSS data (big diamond with error bars).
The nominal fitting parameters for our early- and late-type galaxy subsamples
are also marked in the plot with letters ``E'' and ``L'', respectively.
{\it Lower panel - } The total luminosity distribution is split into the three
main components of early- (E+S0), late-type (Sa-Im) systems, and in the dSph 
component as well, as labelled. Note that the bright 
tail of the galaxy population actually consists of a balanced mix of standard 
spirals and ellipticals while dEs and dSph prevail at the faint-end tail of the 
magnitude distribution. For the sake of reference, the nominal Schechter fit 
for each of these subsamples, according to the previous STY procedure, is reported 
in the upper panel, as labelled.
}
\label{f7}
\end{figure}

The distinctive features about the group galaxy population more clearly emerge as 
far as the whole population of 141 {\it bona fide} member galaxies is considered 
to build up the group luminosity function, as displayed in Fig.~\ref{f7}. 
The tick solid line in the upper panel of the figure represents the global
\citet{schechter76} fit derived with the STY method \citep{sandage79}.
This leads to a value for the estimated parameters of
\begin{equation}
[M^*_g, \alpha]  = [-20.00^{+0.39}_{-0.52}, -0.91^{+0.06}_{-0.07}].
\end{equation}
As a guideline, the nominal fits to the early- and late-type galaxy populations, 
as well as to the dSph component are also added to the panel, although within 
a much larger formal uncertainty, given the lower sample size.

The Schechter parameters derived for the total sample
are well in agreement with the results of \citet{zandivarez10}, who derived the
luminosity function of galaxies in groups from the SDSS Data Release 7.
In their work, these authors report the multicolour Schechter parameters for groups
of different mass. After converting their $^{0.1}g_{AB}$ SDSS magnitudes in 
the appropriate mass range to our magnitude system, according to \citet{blanton07},
and rescaling consistently to our $H_0$ value, we obtain  
\begin{equation}
[M^*_g, \alpha]_{\rm SDSS} = [-20.22\pm 0.06, -0.97\pm 0.05],
\end{equation}
in fairly good agreement (within $1\sigma$) with our result. 
The main contribution to the bright end of the luminosity function
comes from standard spirals and ellipticals, while dEs and dSphs
dominate the faint-end tail.

\begin{figure}
\centerline{
\includegraphics[width=\hsize,clip=]{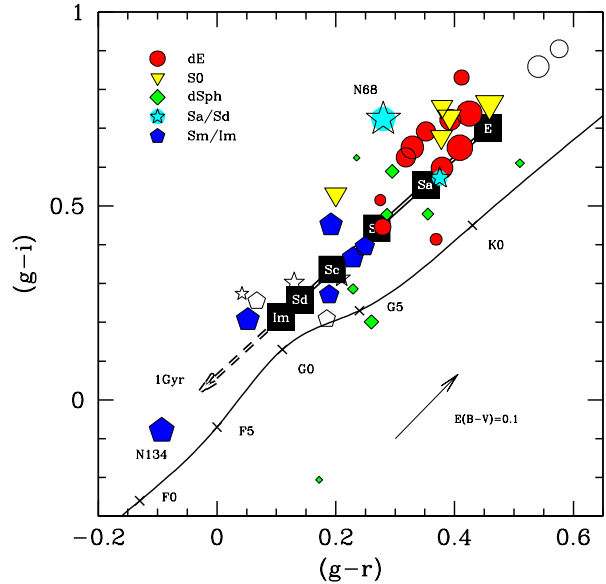}
}
\caption{
The $(g-r)$ vs. $(g-i)$ colour diagram for the 40 galaxies in the {\sc Eso} sample.
Likely group members (``$m_c$'' class 1 and 2 of Table~\ref{a2} and \ref{a3}) have been
singled out as solid markers. Big solid squares indicate the expected colours
for the 15~Gyr \citet{buzzoni05} template galaxy models, as labelled. For the Im model,
the dashed arrow traces the back-in-time colour evolution up to an age of 1~Gyr.
The solid curve marks the reference locus for stars of different spectral type, 
as labelled, according to \citet{straizys72}.
Data have not been corrected for Galactic reddening; the reddening vector
is however reported bottom right in the plot. See text for a full discussion.
Representative error bars for the data are of the order of the biggest marker size. 
}
\label{f8}
\end{figure}

\begin{table*}
\caption{Theoretical colours and star formation parameters for the \citet{buzzoni05} template galaxy models$^{(a)}$}
\label{t4}
\scriptsize
\begin{tabular}{cccccrrr}
\hline\hline
Morph. & $(B-V)$ & $(B-K)$ & $(g-r)$ & b$^{(b)}$ & M/L(B) & M/L(V) & M/L(g)  \\
Type         &         &   &      &  & \multicolumn{3}{c}{\hrulefill~~[M$_\odot$/L$_\odot$]~~\hrulefill} \\
\hline
E~~& 0.919 & 4.140 & 0.456 & 0.0 & 17.87 & 14.08 & 10.17 \\  
S0 & 0.908 & 4.077 & 0.444 & 0.0 & 16.47 & 13.11 &  9.49 \\   
Sa & 0.762 & 3.774 & 0.351 & 0.2 & 10.08 &  9.18 &  6.82 \\   
Sb & 0.649 & 3.455 & 0.269 & 0.5 &  6.07 &  6.13 &  4.66 \\   
Sc & 0.555 & 3.188 & 0.194 & 0.9 &  3.33 &  3.67 &  2.84 \\   
Sd & 0.493 & 2.990 & 0.140 & 1.3 &  1.77 &  2.06 &  1.62 \\   
Im & 0.457 & 2.876 & 0.108 & 1.8 &  1.08 &  1.30 &  1.03 \\   
\hline
\multicolumn{8}{l}{$^{(a)}$ At the reference age of 15 Gyr}\\
\multicolumn{8}{l}{$^{(b)}$ Birthrate, $b = {\rm SFR}_0/\langle {\rm SFR} \rangle$.}\\
\end{tabular}
\end{table*}

An additional view of the distinctive spectrophotometric properties of the
galaxy population, although restrained to the 40 galaxies of the {\sc Eso} sample, can be gained 
through a two-colour diagnostic, like the $(g-r)$ vs.\ $(g-i)$ diagram of Fig.~\ref{f8}.
The galaxy distribution, according to the different morphological types,
is compared with the \citet{buzzoni05} template galaxy models and with the
empirical stellar locus as derived from the Vilnius spectral catalog
\citep{straizys72}. Again, only likely members are plotted with solid
markers. As expected, galaxies univocally discriminate with respect to
stars displaying a ``redder'' $(g-i)$ colour for fixed value of
$(g-r)$ (due to the red-giant stellar component contributing to the galaxy SED).
Early-type galaxies (both dEs and S0s) fairly well match
the \citet{buzzoni05} theoretical locus for ellipticals, although displaying
slightly bluer colours, as expected by their lower mass compared to
``standard'' systems (the same effect can also be easily recognized in the
histograms of Fig.~\ref{f5}). In a few cases, however, the excursion
toward even bluer colours, more appropriate to intermediate-type spirals,
makes evident some star-formation activity among the dEs of lowest mass.
Direct evidence in this sense has been collected for N50 \citep{cb01}
displaying several blue knotty regions around its core. 
This may also be the case for N55, N93C, and N138, among the faintest dwarf
ellipticals easily recognized also in  Fig.~\ref{f6} as 
``embedded'' in the dSph region of the c-m diagram. Similarly, the case of N42, 
a relatively bright S0, seems worth of special attention for its exceedingly blue colours 
(see next section on this regard, and Cellone et al. 2011, in preparation,
for an in-depth spectral analysis of this galaxy).

As far as the ``magellanic'' galaxy component (namely types Sm/Im) is concerned, 
one has to report its large colour spread in Fig.~\ref{f8}, a feature that 
clearly calls for a wide range of star-formation scenarios. 
The striking case of galaxy N134, the bluest and brightest ($M(g) \sim -17.5$) object 
within this group, is worth of mention in this regard. Its colours are consistent
with those of F-type stars, a feature that calls for a very young ($t \ll 1$~Gyr)
age, when compared with template galaxy models of this morphology.
N134 lies at 2.6~arcmin ($\sim 25$~kpc) projected distance from the centre of NGC~5054,
partially overlapping the north arm of this luminous three-armed spiral.
One could guess that the likely interaction between both galaxies may be the
trigger for the intense star forming activity in N134; whether or not this interaction
is also responsible for the peculiar arm morphology of NGC~5054 may still be matter of 
debate \citep[e.g.][]{sandage94,eskridge02}.

In spite of the fairly bunched distribution in the c-m diagram, the
population of dSphs stands out in Fig.~\ref{f8} for its extreme colour spread,
that largely encompasses the full theoretical locus for the different
morphological types. Part of this peculiar distribution has certainly
to be advocated to the large photometric uncertainties, given the
very faint magnitude of these galaxies. The low mass of these systems also
makes the integrated colours to be more easily biased by a few outstanding stars 
inside each galaxy, consequent to fresh star formation. This may also give reason, 
for instance, of the bridging distribution of these galaxies across the stellar locus, 
as evident in Fig.~\ref{f8}.
In any case, it is once more confirmed the puzzling essence of dwarf spheroidals, 
as multifaceted stellar systems able to easily change their 
distinctive look such as to escape any obvious classification within the 
reference evolutionary framework.

\subsection{Inferred mass distribution and stellar birthrate}

As a further step in our analysis, one could take advantage of the accurate distance 
to the group \citep{cb05}, and a full morphological and photometric characterization 
of its galaxy members, to try a direct estimate of galaxy stellar mass and therefrom 
assess other fundamental physical properties of the group as a whole, such as
its total bright mass and current stellar birthrate.

This can be done by matching our observed galaxy magnitudes with the appropriate stellar 
$M/L$ ratio, as proposed by theoretical models. For its better accuracy, $g_{27}$ and
$V_{27}$ photometry has been taken as a reference for the {\sc Eso} and {\sc Casleo} samples, 
respectively. Again, the \citet{buzzoni05} models provide the set of relevant quantities, 
as summarized in Table~\ref{t4}. As we previously discussed, the dSph case is
hard to fit within the canonical scenario of standard morphological types. However,
according to the observed colour distribution for these galaxies (see, again 
Fig.~\ref{f5} and \ref{f8}) one could envisage a composite stellar population 
vaguely resembling that of Sb spirals.

To consistently scale Gunn magnitudes to solar units we assumed for the Sun a colour 
$(V-g)_\odot = -0.14$ obtained by convolving a \citep{straizys72} spectrum for a G2V 
star with the relevant $V$ and $g$ filters. Recalling, in addition, that 
$M(V)_\odot = +4.79$ and $M(B)_\odot = +5.45$ \citep{portinari04}, then 
$M(g)_\odot = +4.93$. Apparent $g$ magnitudes for our {\sc Eso}
galaxies eventually translate into ${\cal L}_g$ solar luminosities as
\begin{equation}
{\cal L}_g = 10^{-0.4\,[(g-32.58) -4.93)]},
\end{equation}
while, for the {\sc Casleo} $V$ magnitudes, we have
\begin{equation}
{\cal L}_V = 10^{-0.4\,[(V-32.58) -4.79)]}.
\end{equation}
If one wants to further complete the sample with the supplementary {\it bona fide}
member galaxies comprised in the \citet{fs90} and \citet{mendel08} catalogs,
then from their $B$ photometry we can write,
\begin{equation}
{\cal L}_B = 10^{-0.4\,[(B-32.58) -5.45)]}.
\end{equation}
Galaxy bright mass simply follows either as
\begin{equation}
M^*_{\rm gal} = \left\{
\begin{array}{l}
\left({M\over L_g}\right)\,{\cal L}_g\\
\\
\left({M\over L_V}\right)\,{\cal L}_V\\
\\
\left({M\over L_B}\right)\,{\cal L}_B\\
\end{array}
\right.
\label{eq:mstar}
\end{equation}
providing to chose the appropriate $M/L$ ratio from Table~\ref{t4}
according to galaxy morphology.

\begin{table}
%\centering
\caption{Inferred stellar mass for the present galaxy sample$^{(a)}$}
\label{t5}
\scriptsize
\begin{tabular}{lrrclrr}
\hline\hline
\multicolumn{3}{c}{\hrulefill~ESO sample~\hrulefill} & & \multicolumn{3}{c}{\hrulefill~{\sc Casleo} sample~\hrulefill} \\
Name & T & $\log M^*_{\rm gal}$ & & Name & T & $\log M^*_{\rm gal}$ \\
% & & [M$_\odot$] & [M$_\odot$~yr$^{-1}$]  &  & & [M$_\odot$] & [M$_\odot$~yr$^{-1}$] \\
\hline
N17  &  0 &  9.76 &  &   N18   &  1  & 10.68   \\
N20  & -5 &  8.59 &  &   N19   &  9  &  8.25   \\
N24  &  9 &  8.25 &  &   N29   & -5  & 10.17   \\
N30  & -5 &  9.90 &  &   N38   & -5  &  9.43   \\
N31  & 10 &  8.49 &  &   N46A  & -4  &  7.88   \\
N32  &  0 & 10.71 &  &   N49A  & -4  &  7.66   \\
N34  & -5 &  9.54 &  &   N51   &  0  &  9.78   \\
N42  &  0 &  9.90 &  &   N53   & -5  &  9.70   \\
N49  & 10 &  8.72 &  &   N57   & -5  &  8.81   \\
N50  & -5 &  9.92 &  &   N58   & 10  &  7.87   \\
N54  & 10 &  8.69 &  &   N61   & -4  &  7.97   \\
N54A & -4 &  7.13 &  &   N63   &  0  & 10.47   \\
N55  & -5 &  8.15 &  &   N66   & -5  &  9.48   \\
N56  & -4 &  8.13 &  &   N72A  & -4  &  7.86   \\
N62  & -4 &  7.81 &  &   N76   & -5  &  9.84   \\
N64A & -4 &  7.86 &  &   N79   &  0  &  9.85   \\
N68  &  2 & 10.99 &  &   N80   & -5  &  8.40   \\
N70  & -5 &  9.12 &  &   N82   &  0  & 10.45   \\
N70A & -4 &  7.40 &  &   N82A  & -4  &  7.52   \\
N71  & -5 &  9.44 &  &   N94   &  0  & 10.15   \\
N75  &  0 &  9.85 &  &   N94A  & -4  &  7.73   \\
N83  & -5 &  9.34 &  &   N94B  & -4  &  7.88   \\
N83A & -4 &  7.65 &  &   N95   & -5  &  9.07   \\
N89  & -5 &  9.11 &  &   N95A  & -4  &  8.47   \\
N93A & -4 &  8.09 &  &   N108  & -5  &  9.34   \\
N93B & -4 &  7.09 &  &   N116  & -5  &  8.57   \\
N93C & -5 &  8.05 &  &   N117  &  0  & 10.12   \\
N134 &  9 &  9.01 &  &   N122  & -5  &  9.31   \\
N138 & -5 &  8.80 &  &   N131  &  0  & 10.14   \\
N153 &  0 & 10.07 &  &   N131A & -4  &  8.19   \\
N155 &  2 &  9.37 &  &   N149  & 10  &  8.24   \\
N156 & 10 &  8.25 &  &	       &     &         \\
\hline\hline
\noalign{$^{(a)}$ Only likely member galaxies (code $mc \le 2$ in Table~\ref{a2} and \ref{a3})
are considered. Galaxy stellar mass $M^*_{\rm gal}$ in solar units from
the assumed $M/L$ ratio and absolute $g$ magnitude, according to
eq.~(\ref{eq:mstar}).}
\end{tabular}
\end{table}

\begin{figure}
\centerline{
\includegraphics[width=0.98\hsize,clip=]{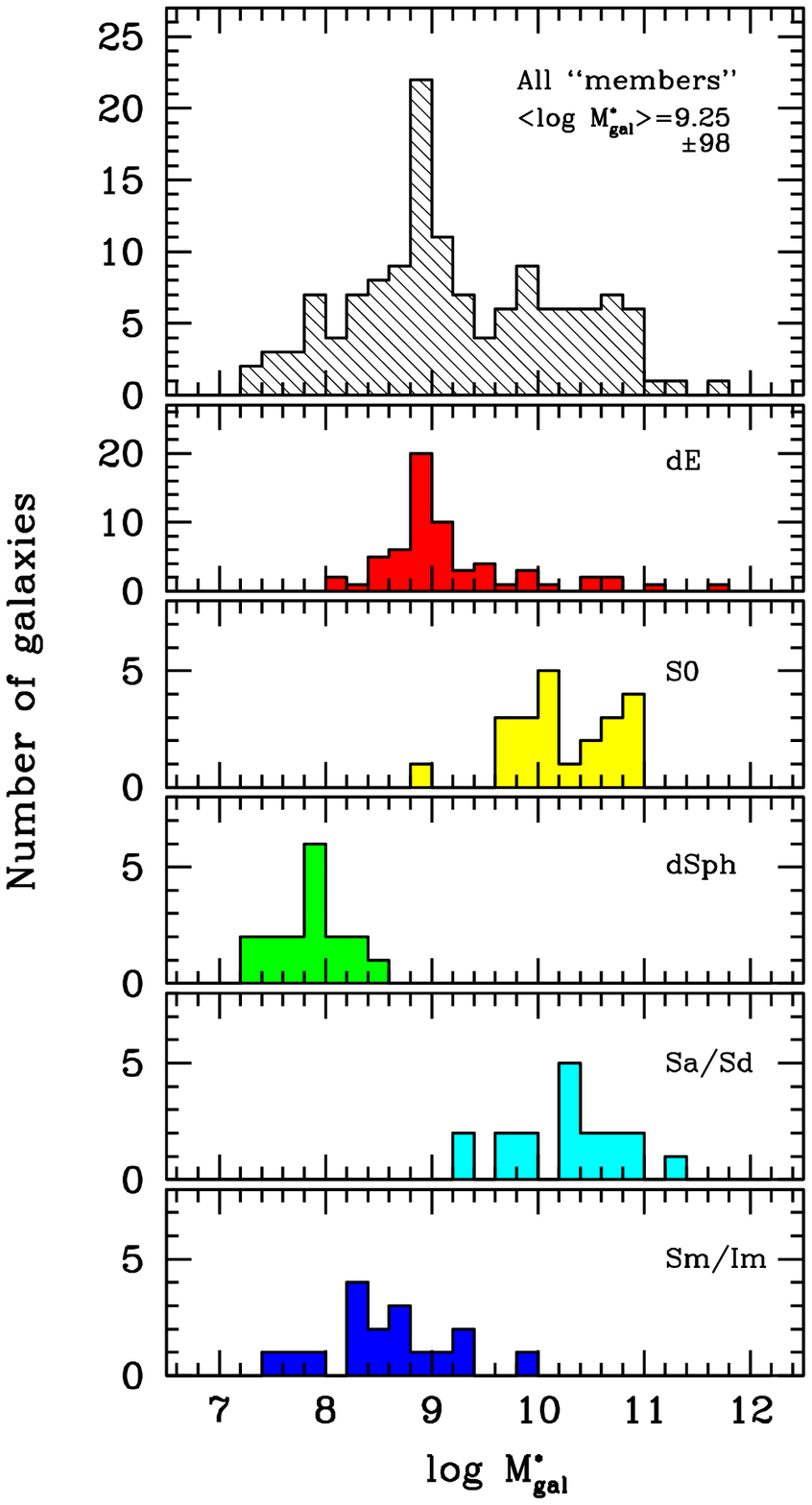}
}
\caption{
The bright stellar mass ($M^*_{\rm gal}$) of the 136 likely member galaxies
with available photometry and morphological classification
(63 from our sample, 50 more from the \citet{fs90} and 23 from the \citet{mendel08}
catalogs), as inferred from the 
absolute $g$, $V$ or $B$ luminosity and the theoretical $M/L$ ratio as from the 
\citet{buzzoni05} template galaxy models of Table~\ref{t4}. A mean logarithmic
mass of $\langle \log M^*_{\rm gal}\rangle = 9.25 \pm 0.98$, or 
$\langle M^*_{\rm gal}\rangle = 1.8\,10^9$~M$_\odot$ is derived for the whole
sample, as reported in the top panel.
The different morphological groups are singled out along the different panels, 
as labelled. Note a clear mass segregation effect, with systematically
more massive dE's and dS0's galaxies compared to very low-mass dSph's and Im's.
}
\label{f9}
\end{figure}

\begin{table*}
%\centering
\caption{Inferred stellar mass for the supplementary
set of confirmed member galaxies in the \citet{fs90} and \citet{mendel08} catalogs}
\label{t6}
\scriptsize
\begin{tabular}{lrrrclrrrr}
\hline\hline 
\multicolumn{5}{c}{Confirmed ($m_c \le 2$) members} & \multicolumn{5}{c}{Confirmed members} \\                                       
\multicolumn{5}{c}{in \citet{fs90}} & \multicolumn{5}{c}{in \citet{mendel08}$^{(a)}$} \\                                       
\hline
Name & T & B & $\log M^*_{\rm gal}$ & $\qquad$ &  Name & T & B$^{(b)}$ & K & $\log M^*_{\rm gal}$ \\
\hline                                  	     
N1   & -5 & 18.4 &  9.10  & $\qquad$ &   N3			&   3& 14.3  &  9.34 &  10.28          \\
N5   & -5 & 13.2 & 11.18  & $\qquad$ &   N45			&  10& 18.9  &    \ldots &  7.69       \\
N7   & -5 & 19.2 &  8.78  & $\qquad$ &   N69			&  -5& 18.7  &    \ldots &  8.98       \\
N9   &  0 & 14.5 & 10.63  & $\qquad$ &   N96			&  \ldots& 18.5  &    \ldots &  \ldots \\
N11  &  0 & 18.9 &  8.87  & $\qquad$ &   N146			&  -5& 19.1  &    \ldots &   8.82      \\
N15  & -5 & 14.6 & 10.62  & $\qquad$ &   N158			&   0& 16.9  &  13.15 &   9.67         \\
N16  & 10 & 19.4 &  7.49  & $\qquad$ &   2MASX:J13085477-1636106  &  \ldots& 16.37 &  12.97 &  \ldots    \\
N22  & -5 & 18.9 &  8.90  & $\qquad$ &   2MASX:J13094408-1636077  &   0& 13.87 &  9.75 &  10.88          \\
N25  & -5 & 19.0 &  8.86  & $\qquad$ &   2MASX:J13091671-1653115  &  \ldots& 16.40 &  12.83 &  \ldots    \\
N26  & -5 & 18.9 &  8.90  & $\qquad$ &   2MASX:J13095347-1631018  &   0&  {\it 15.6:}  &  11.52 &  10.19 \\
N27  &  4 & 14.2 & 10.32  & $\qquad$ &   2MASX:J13100952-1616458  &  \ldots& 17.23 &  12.34 &  \ldots    \\
N36  & -5 & 19.1 &  8.82  & $\qquad$ &   2MASX:J13102493-1655578  &  \ldots& 13.94 &  10.11 &  \ldots    \\
N40  &  5 & 16.3 &  9.21  & $\qquad$ &   2MASX:J13114576-1915421  &   1&  {\it 13.2:}  & 9.44 &   10.94  \\
N41  & -5 & 18.8 &  8.94  & $\qquad$ &   2MASX:J13114703-1847312  &   4& 15.67 &  12.93 &   9.73         \\
N48  & -5 & 19.7 &  8.58  & $\qquad$ &   2MASX:J13115849-1644541  &  \ldots& 15.29 &  11.96 &  \ldots    \\
N60  & -5 & 18.9 &  8.90  & $\qquad$ &   2MASX:J13123543-1732326  &   5& 13.14 &   8.85 &   10.48        \\
N64  &  1 & 13.9 & 10.66  & $\qquad$ &   2MASX:J13133433-1525554  &   8& 13.71 &  10.85 &  9.98          \\
N67  & -5 & 19.3 &  8.74  & $\qquad$ &   2MASX:J13143041-1732009  &  \ldots& 15.23 &  11.54 &  \ldots    \\
N72  &  0 & 13.9 & 10.87  & $\qquad$ &   2MASX:J13151278-1758006  &   7& 14.61 &  12.32 &   9.62         \\
N74  & -5 & 19.4 &  8.70  & $\qquad$ &   2MASX:J13153736-1452209  &  \ldots&  \ldots  &  11.24 &  \ldots \\
N77  & -5 & 19.7 &  8.58  & $\qquad$ &   2MASX:J13164875-1620397  &  \ldots& 15.45 &  11.60 &  \ldots    \\
N78  & -5 & 15.0 & 10.46  & $\qquad$ &   2MASX:J13165533-1756417  &   0& 15.26 &  11.94 &  10.32         \\
N81  & -5 & 19.1 &  8.82  & $\qquad$ &	MASX:J13171239-1715162  &   4& 14.24 &  10.24 &   10.30   \\
N84$^{(c)}$ & -5 & 11.9 & 11.70 & $\qquad$ & 2MASX:J13182685-1545599  &   3&  {\it 15.0:}  &  11.57 &  \\
N87  & -5 & 19.1 &  8.82  & $\qquad$ &	  2MASX:J13183034-1436319  &   5& {\it 13.7:} &  10.48 & 10.25 \\
N91  & -5 & 19.8 &  8.54  & $\qquad$ &	  2MASX:J13184125-1904476  &   1& 15.77 &  12.06 &   9.91      \\
N97   & -5 & 19.0 &  8.86 & $\qquad$ &     2MASX:J13185909-1835167  &	0& 14.14 &  10.12 &   10.77    \\
N100  & -5 & 14.3 & 10.74 & $\qquad$ &     2MASX:J13191752-1509252  &  \ldots&  \ldots  & 10.85 & \ldots \\ 
N102  &  0 & 13.7 & 10.95 & $\qquad$ &     2MASX:J13192062-1450402  &	5& 13.14 &   9.16 &   10.48   \\
N103  & -5 & 19.2 &  8.78 & $\qquad$ &     2MASX:J13192221-1509232  &  \ldots&  \ldots  &  13.16 &  \ldots \\ 
N104  & -5 & 18.9 &  8.90 & $\qquad$ &     2MASX:J13201698-1448455  &  \ldots&  \ldots  &  11.33 &  \ldots \\
N105  & -5 & 19.0 &  8.86 & $\qquad$ &     2MASXi:J1320185-163215   &  \ldots& 14.44 &  11.13 &  \ldots    \\
N107  &  0 & 13.8 & 10.91 & $\qquad$ &     6dF:j1311150-180610      &  \ldots& 16.16 &  13.80 &  \ldots    \\
N112  & -5 & 18.8 &  8.94 & $\qquad$ &     6dF:j1313501-173048      &  \ldots& 16.84 &    \ldots &  \ldots \\
N113  & -5 & 19.0 &  8.86 & $\qquad$ &     GEMS\_N5044\_5	    &  \ldots&  \ldots  &    \ldots &  \ldots	\\ 
N121  & -5 & 18.6 &  9.02 & $\qquad$ &     GEMS\_N5044\_14	    &  \ldots&  \ldots  &    \ldots &  \ldots	 \\
N123  & -5 & 18.3 &  9.14 & $\qquad$ &     GEMS\_N5044\_7	    &  \ldots& 17.68 &  13.62 &  \ldots    \\
N126  & -5 & 19.1 &  8.82 & $\qquad$ &     GEMS\_N5044\_18	    &  \ldots&  \ldots  &    \ldots &  \ldots \\	  
N127  & -5 & 18.6 &  9.02 & $\qquad$ &     GEMS\_N5044\_1	    &  \ldots&  \ldots  &    \ldots &  \ldots \\	  
N128  & -5 & 19.0 &  8.86 & $\qquad$ &     PGC:045257		    &	9& 15.05 &    \ldots &  9.23   \\
N133  & -5 & 18.6 &  9.02 & $\qquad$ &     PGC:046242		    &  \ldots& 16.96 &    \ldots &  \ldots \\  
N135  & -5 & 18.6 &  9.02 & $\qquad$ &     PGC:046402		    &	9& 16.31 &    \ldots &    8.72 \\
N137$^{(d)}$ &  3 & 11.5 & 11.40 & $\qquad$ &  PGC:046494  	     &   8& 15.54 &    \ldots &   9.24 \\
N141  &  5 & 13.5 & 10.33 & $\qquad$ &     HIPASS:J1312-15	    &  \ldots&  \ldots  &    \ldots &  \ldots \\	 
N142  & -5 & 18.5 &  9.06 & $\qquad$ &     HIPASS:J1320-14	    &  \ldots&  \ldots  &    \ldots &  \ldots \\	  
N144  & -5 & 14.8 & 10.54 & $\qquad$ &     FGC1563		    &	8& 16.4  &  13.23 &   8.90  \\
N147  & -5 & 19.2 &  8.78 & $\qquad$ &   &  &  &  & \\
N151  & -5 & 19.0 &  8.86 & $\qquad$ &   &  &  &  & \\
N154  & -5 & 19.4 &  8.70 & $\qquad$ &   &  &  &  & \\
N162  &  9 & 16.9 &  8.49 & $\qquad$ &   &  &  &  & \\
\hline\hline
\multicolumn{10}{l}{$^{(a)}$ This includes six entries originally 
classified as $m_c = 3$ by \citet{fs90}, whose membership}\\
\multicolumn{10}{l}{ has been spectroscopically confirmed, on the contrary, by \citet{mendel08}.\hfill} \\
\multicolumn{10}{l}{$^{(b)}$ Entries in italics estimated from $K$ magnitudes, according to the 
reference $(B-K)$ colour of}\\
\multicolumn{10}{l}{theoretical templates from Table~\ref{t4}}\\
\multicolumn{10}{l}{$^{(c)}$ N84 $\equiv$ NGC~5044.}\\
\multicolumn{10}{l}{$^{(d)}$ N137 $\equiv$ NGC~5054.}\\
\multicolumn{10}{l}{Galaxy stellar mass $M^*_{\rm gal}$ in solar unit from
the assumed $M/L$ ratio and absolute $B$ (or, in lack, $K$) magnitude,}\\
\multicolumn{10}{l}{ according to eq.~(\ref{eq:mstar}) and Table~\ref{t4} colours.\hfill}\\
\end{tabular}
\end{table*}

The results of our excercise, for the 63 likely member galaxies (membership code $m_c \le 2$
in Tables~\ref{a2} and \ref{a3}) in our sample plus the \citet{fs90} and \citet{mendel08}
extra sample (50+23 objects with available photometry and morphological classification)
are reported in Table~\ref{t5}, and \ref{t6}, respectively, where 
for each entry the nominal output of eq.~(\ref{eq:mstar}) is 
displayed.\footnote{For those \citet{mendel08} galaxies lacking $B$ photometry,
eq.~(\ref{eq:mstar}) has still been used, relying on the observed $K$ magnitude
and an appropriate $(B-K)$ colour, according to morphology, as from Table~\ref{t4}.}
The $M^*_{\rm gal}$ distribution for the likely-member galaxy sample is displayed in Fig.~\ref{f9}
splitting the contribution of the different morphological classes in the vertical
panel sequence.
A sharp lower-mass cutoff is evident in the distribution with a definite lack of galaxies 
below $10^7$~M$_\odot$ and with the $M^*_{\rm gal} \lesssim 10^8$~M$_\odot$
population almost uniquely consisting of dSph type. The mean
logarithmic mass of the whole sample results $\langle \log M^*_{\rm gal}\rangle = 9.25 \pm 0.98$,
that is $\langle M^*_{\rm gal}\rangle = 1.8\,10^9$~M$_\odot$, with galaxy
mass spreading along nearly one order of magnitude.

A comparison with \citet{mendel09} can be attempted, by relying on the mass estimate 
for the 28 galaxies in common. For this set, these authors obtain
a total of $1.0\,10^{11}$~M$_\odot$ versus our estimate of $3.8\,10^{11}$~M$_\odot$, 
that is roughly a factor of four larger. This difference is mainly due to the adopted M/L 
ratio which, for \citet{mendel09}, comes from a plain SSP fitting of each galaxy, 
based on the \citet{bruzual03} population synthesis models.
As stated by the authors, their procedure disregarded morphology details and star 
formation history, thus forcedly leading to younger ``luminosity-weighted''
ages for their galaxies (see \citealp{buzzoni11} and eq.~12 in \citealp{buzzoni05}
for an estimate of this effect). 
In addition, the correspondingly lower M/L figures would be further decreased by 
the adopted \citet{chabrier03} IMF. Compared to the Salpeter case, in fact, a 
\citet{chabrier03} dwarf-depleted IMF leads to a brighter SSP (and therefore 
to a lower M/L ratio) for fixed total stellar mass.
According to our calculations, the total bright mass stored in the member galaxies of 
the group amounts to
$M^*_{\rm gal}({\rm tot}) = \sum M^*_{\rm gal}  = 2.3~10^{12}$~M$_\odot$, 
almost half of which being comprised in the three brightest members, that is
NGC~5044 (22\%), 5054 (11\%) and the gorgeous Sab galaxy N68 (4\%) (see CB05). 

Based on the template galaxy models of Table~\ref{t4}, a nominal estimate of the 
current galaxy SFR for group members can even be guessed, through the model birthrate, 
i.e.\ $b = {\rm SFR}_0 /\langle {\rm SFR} \rangle$, 
so that
\begin{equation}
{\rm SFR}_0  = {{b\,{M^*_{\rm gal}}}\over {t_{\rm gal}}}, 
\label{eq:sfr}
\end{equation}
being $t_{\rm gal}$ the assumed galaxy age.
By summing up all the entries of Table~\ref{t5} and \ref{t6} we have that fresh stars 
are produced inside the group at a rate of roughly 23~M$_\odot$\,yr$^{-1}$. Assuming 
galaxies to be one Hubble time old, in force of eq.~(\ref{eq:sfr}) this leads to a global birthrate for the group of
$b_{\rm tot} \simeq 0.15$. 

\begin{figure}
\centerline{
\includegraphics[width=\hsize,clip=]{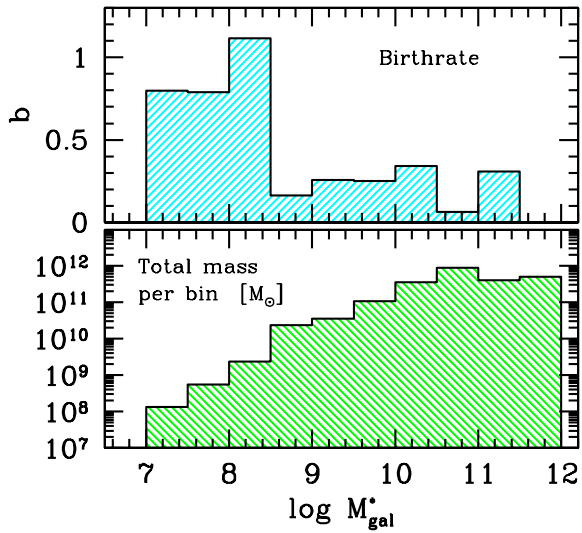}
}
\caption{
{\it Upper panel:} Mean birthrate versus galaxy stellar mass, as inferred from
the match with the \citet{buzzoni05} template galaxy models (see text for
the detailed procedure). The total sample of the 136 likely-member galaxies
of Fig.~\ref{f9} is considered. Note the ``downsizing'' effect, with low-mass
galaxies (of prevailing dSph/Im morphological type) supplying the fresh
star formation within the NGC~5044 group. Fresh stars are produced 
inside the group at a rate of roughly 23~M$_\odot$\,yr$^{-1}$, while bright
stellar mass sampled by our observations amounts to  
$M^*_{\rm gal}({\rm tot}) = \sum M^*_{\rm gal}  = 2.3~10^{12}$~M$_\odot$.
This total is singled out in the {\it lower panel} in terms of the
cumulative galaxy mass fraction for each mass bin.
}
\label{f10}
\end{figure}

To much extent, such a low figure is mostly induced
by the prevailing contribution of  early-type galaxies among the more massive 
members of the NGC~5044 group (see the disaggregated distribution of Fig.~\ref{f9}).  
As a result, an important fraction of $M^*_{\rm gal}({\rm tot})$ resides in nearly quiescent
stellar aggregates (see Fig.~\ref{f10}), while only very low-mass systems 
(mostly dSph and Im types) are still able to feed fresh stars to the global environment.
Even at this small mass scale, therefore, the emerging picture seems consistent with 
the so-called down-sizing mechanism \citep{cowie96,gavazzi93}, that is an inverse 
dependence of the birthrate with galaxy mass as the imposing paradigm for galaxy
formation in the Universe.

\section{Spectroscopic properties}

A preliminary analysis of the spectroscopic material collected along our survey 
has already been provided in \citet[][see Table~1 therein]{cb05}, where redshift 
measurements have been presented for 14 galaxies of the sample. For a total of 10 
objects, the paper confirmed group membership, while 4 galaxies resulted to lie 
in the background. After thorough reconsideration of the data, also including
the {\sc Casleo} observing sessions, we can further expand 
here the original output by including 7 more objects, for which a confident 
redshift measure can be obtained together with a rough morphological classification 
according to the detected spectral features. For a further set of 3 spectra no 
definite conclusions can be achieved, mostly due to the extremely poor S/N level.
Our spectroscopic results, then, extend and complement the works of \citet{mendel08,mendel09},
which provided 103 (mostly new) redshifts for galaxies in the NGC~5044 Group catalog,
as well as a stellar population analysis through Lick indices for a subset of 67 group
members.

\begin{table}
%\centering
\caption{Spectroscopic redshift for observed galaxies}
\label{t7}
\scriptsize
\begin{tabular}{lccl}
\hline\hline
\multicolumn{4}{c}{\hrulefill~~Members~~\hrulefill}\\
ID & Type & cz & Remarks\\
   &      & [km\,s$^{-1}$] &  \\
\hline
N17  &  0 & 2682 & H$\beta$ emission \\ 
N29  & -5 & 2351 & from {\sc Casleo}  \\
N30  & -5 & 2411 &              \\
N34  & -5 & 2661 &              \\
N42  &  0 & 2462 &              \\
N49  & 10 & 1499 &              \\
N50  & -5 & 2392 &  H$\beta$ emission \\
N75  &  0 & 1831 &              \\
N84  & -5 & 2710 &  NGC 5044  - H$\beta$ emission \\
N153 &  0 & 2816 &              \\
N155 &  2 & 2922 &              \\
\hline
\multicolumn{4}{c}{\hrulefill~~Background~~\hrulefill} \\
ID   & Type & z  & Remarks \\ 
\hline
N109 &  8 & 0.018 &             \\
N33  &  5 & 0.045 &             \\ 
N152 &  7 & 0.045 & H$\beta$+ [O{\sc iii}]$_{5007}$ emission \\
N39  & -5 & 0.091 &             \\
B1   & {\it -5?} & 0.097 &             \\
B3   & -5 & 0.096 & H$\beta$+ [O{\sc iii}]$_{5007}$ emission \\
B2   & {\it -5?} & 0.282 &             \\
B4   & $>${\it +5?} & 0.277 & [O{\sc ii}]$_{3727}$ + H$\beta$ emission \\
B6   &  {\it -5?} & 0.358 &             \\
B5   &  {\it -5?} & 0.424 &             \\
\hline
\multicolumn{4}{c}{\hrulefill~~Unclassified~~\hrulefill} \\
N20  & -5 &  ?    & [O{\sc iii}]$_{5007}$ emission? \\
N55  & -5 &  ?    &             \\
N156 & 10 &  ?    &             \\
\hline\hline
\noalign{Galaxy type, in italics, refer to a spectroscopic classification alone.}
\end{tabular}
\end{table}

The summary of our results is presented in Table~\ref{t7}, which also recollects
the data of \citet{cb05} for reader's better convenience. Out of the 25 objects
considered by the observations, 10 galaxies, plus NGC~5044, belong to the same
physical group, while behind one may guess the presence of at least three 
galaxy aggregations, that also match the \citet{mendel08} data, located respectively 
at $z \sim 0.045, 0.095$, and 0.28, the farthest one actually confirmed by the 
coherent X-ray emission studied by XMM-Newton \citep{gastaldello07}.
The most distant galaxy in our sample is object B5, located at $z = 0.42$.
According to Table~\ref{t7}, a roughly similar fraction of member and background 
galaxies is not a surprising feature, of course, given our observing strategy, that
included a ``bonus'' object for each target pointing, as explained in Sec.~2.2. 
In this regard we stress the very high reliability of morphological membership assignment 
by \citet{fs90}, which reaches $\sim 91$\% for $m_c=1$ galaxies \citep{mendel08}.
As a reference for possible future investigations, we report in Table~\ref{a4} 
the accurate coordinates for the 6 ``B'' galaxies catched
by the spectrograph slit\footnote{Note that these coordinates update and complete
the corresponding table originally proposed in \citet{cb05}.}

\begin{figure}
\centerline{
\includegraphics[width=0.86\hsize,clip=]{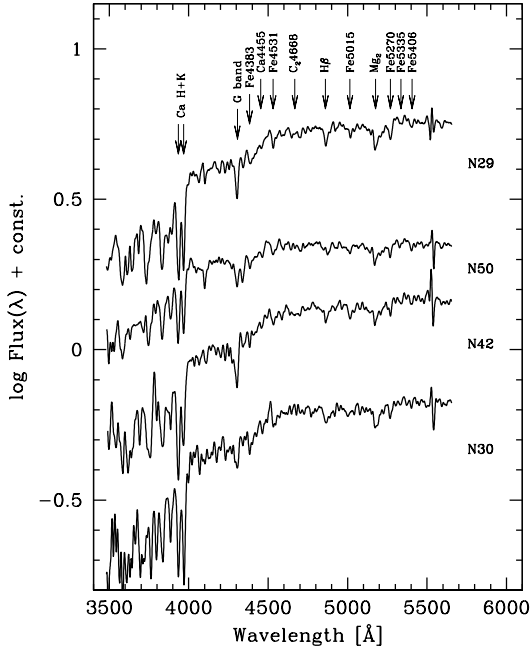}
}
\caption{
Spectral energy distribution for the four galaxies observed at {\sc Casleo}.
For better reading, spectra have been degraded here to a resolution of 12~\AA\ FWHM
and corrected for redshift. The Ca H+K absorption lines together with all the main 
spectral features included in the Lick narrow-band index system (see Table~\ref{t8})
are marked on the top. 
}
\label{f11}
\end{figure}

\begin{figure}
\centerline{
\includegraphics[width=0.86\hsize,clip=]{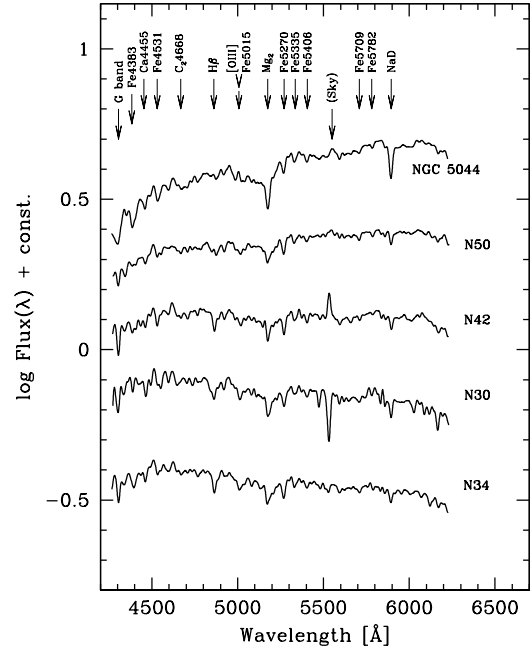}
}
\caption{
Same as Fig.~\ref{f11}, but for a first set of member galaxies observed
at {\sc Eso} including N50, N42 and N30 in common with {\sc Casleo}.
For better reading, spectra have been degraded here to a resolution of 12~\AA\ FWHM
and corrected for redshift. Marked on the top are also all the main 
spectral features included in the Lick narrow-band index system (see Table~\ref{t8}).
}
\label{f12}
\end{figure}

\begin{figure}
\centerline{
\includegraphics[width=0.86\hsize,clip=]{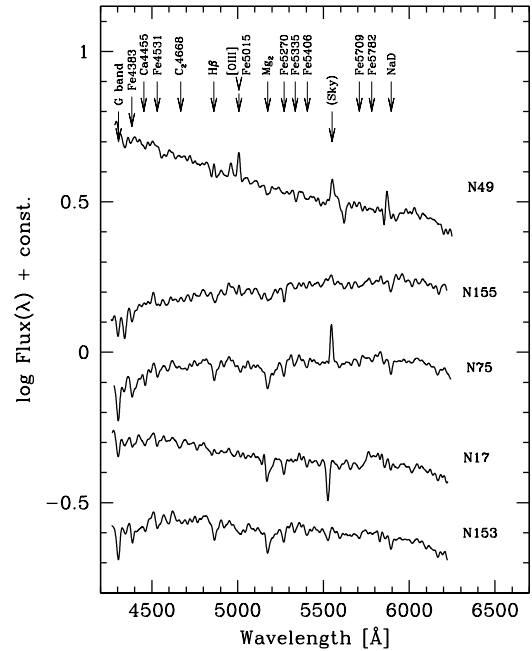}
}
\caption{
Same as Fig.~\ref{f11}, but for a second set of member galaxies observed
at {\sc Eso}. Note the outstanding H$\beta$ and [O{\sc iii}] 5007\AA\ emission
in Im galaxy N49.
For better reading, spectra have been degraded here to a resolution of 12~\AA\ FWHM.
and corrected for redshift. Marked on the top are also all the main 
spectral features included in the Lick narrow-band index system (see Table~\ref{t8}).
}
\label{f13}
\end{figure}

A graphical display of the SED for the 11 member galaxies
is given in the series of Fig.~\ref{f11} to \ref{f13}. The {\sc Casleo} observations,
obtained with a ``bluer'' setup are reported first in Fig.~\ref{f11}, while
the {\sc Eso} spectra are collected in the other two figures. For a closer ``quick-look''
analysis of the two data sets, and to increase S/N in the plots, all the displayed 
spectra have been homogeneously degraded to a FWHM resolution of 12~\AA.
Once comparing the three objects in common between {\sc Eso} and {\sc Casleo} (namely galaxies 
N30, N42, and N50, as in Fig.~\ref{f11} and \ref{f12}) one sees that 
along the wavelength region in common the spectral pattern of absorption features is 
consistently reproduced in both data sets. 

Some difference can be noticed, on the
contrary, for the overall slope of the continuum emission.
One has to consider, in this regard, that {\sc Casleo} spectra have all been taken at 
fixed (E/W) slit position angle; this is not the case
for the {\sc Eso} spectra for which a range of slit inclinations led in 
some cases to a larger departure from the parallactic angle, and therefore to a 
slightly poorer flux calibration. 
This effect has been explored by means of repeated spectra for the dE galaxy N34
taken in different nights. On the basis of the observed drift in the continuum shape
we estimate that the induced internal uncertainty of our flux calibration can be 
quantified in $\sim10$\%.

Concerning the background galaxy population, object B3 is probably the most  
striking case as the object is well resolved in our images, in spite of
its relatively large redshift, and therefore was included in our photometry (see Table~\ref{a2}
and \ref{a3}). Its S\'ersic index ($n=0.22$) corresponds to a profile steeper than
a de~Vaucouleurs law, while the spectrum revealed an active elliptical with strong 
H$\beta$ and [O{\sc iii}]$_{5007}$ line emission.

\begin{table*}
\caption{Lick narrow-band indices}
\label{t8}
\begin{minipage}{21.5cm}
\scriptsize
\begin{tabular}{lrrrrrrrrrrrrrrrrr}
\hline\hline
Name & Type & G    & Fe   & Ca   & Fe   & C$_2$ & H$\beta$ & Fe   & Mg$_1$ & Mg$_2$ & Mgb & Fe   & Fe   & Fe   & Fe   & Fe   & NaD \\
     &      & 4300 & 4383 & 4455 & 4531 & 4668 &          & 5015 &        &        &     & 5270 & 5335 & 5406 & 5709 & 5782 &     \\  
\hline\multicolumn{18}{l}{\underline{ESO sample}} \\
%#    Tipo  g43  fe43  ca44  fe45  fe46    hb  fe50    mg1    mg2   mgb  fe52  fe53  fe54  fe57  fe58   nad
N17  &  0 &  3.22 &  0.78 &  0.60 &  2.16 &  1.88 &  0.07 &  1.22 &  0.016 &  0.138 &  3.44 &  2.27 &  1.77 &  1.48 &       &       &  0.38 \\ 
     &    &       &       &       &       &       &{\it (0.83)} &   &        &        &       &       &       &       &       &       &       \\ 
N30  & -5 &  3.48 &  0.68 &  2.26 &  3.77 &  4.47 &  1.44 &  5.08 &  0.064 &  0.189 &  2.82 &  2.99 &  2.02 &  1.24 &  0.61 &       &  1.78 \\ 
N34  & -5 &  2.47 &  3.37 &  1.66 &  2.44 &  4.23 &  2.34 &  4.91 &  0.054 &  0.166 &  2.73 &  3.19 &  2.02 &  1.18 &  0.71 &  0.72 &  1.85 \\ 
N42  &  0 &  3.36 &  2.63 &  1.27 &  2.99 &  6.00 &  2.26 &  4.53 &  0.051 &  0.136 &  2.35 &  3.22 &  2.59 &  1.36 &  0.55 &  0.48 &  1.48 \\ 
N49  & 10 & -3.44 & -0.82 &  0.25 & -0.22 &  0.60 & -0.29 &  1.12 &  0.001 &  0.072 &  1.89 &  0.39 &  1.43 &  0.95 & -0.48 &  0.06 &  1.95 \\ 
N50  & -5 &  1.43 &  1.57 &  0.98 &  2.63 &  2.17 &  0.54 &  4.50 &  0.041 &  0.130 &  1.97 &  2.97 &  1.90 &  1.03 &  0.79 &  0.78 &  1.76 \\ 
     &    &       &       &       &       &       &{\it (1.68)} &   &        &        &       &       &       &       &       &       &       \\ 
N75  &  0 &  5.08 &  5.06 &  1.58 &  2.45 &  4.37 &  1.95 &  4.88 &  0.055 &  0.171 &  2.69 &  3.13 &  2.54 &  1.57 &  1.00 &  0.58 &  1.92 \\ 
N84  & -5 &  3.72 &  4.51 &  1.07 &  3.36 &  5.84 &  0.04 &  4.14 &  0.160 &  0.286 &  4.60 &  2.83 &  1.94 &  1.29 &  0.54 &  0.66 &  4.76 \\ 
     &    &       &       &       &       &       &{\it (0.64)} &   &        &        &       &       &       &       &       &       &       \\ 
N153 &  0 &  4.97 &  3.26 &  1.36 &  3.40 &  5.20 &  2.06 &  4.76 &  0.036 &  0.161 &  2.84 &  2.67 &  1.87 &  1.62 &  0.73 &  0.27 &  1.63 \\ 
N155 &  2 &  1.88 &  3.55 &  0.09 &  1.99 &  0.93 &  1.35 &  2.76 &  0.034 &  0.090 &  0.79 &  2.45 &  1.40 &  1.06 &  0.67 &  0.56 &  2.14 \\ 
%#
%#    	   	       CAS  LEO (1  0.1 A   FWHM)  	   	   	    	     	     	     	     	     	     	     	     
\\
\multicolumn{18}{l}{\underline{{\sc Casleo} sample}} \\
N29  & -5 &  4.95 &  4.19 &  0.77 &  2.88 &  4.08 &  2.06 &  3.78 &  0.042 &  0.154 &  2.50 &  3.44 &  1.72 &  1.42 &       &       &       \\ 
N30  & -5 &  4.93 &  5.54 &  1.36 &  2.24 &       &  1.57 &  3.14 &  0.040 &  0.138 &  2.59 &  2.21 &  0.45 &  0.80 &       &       &       \\ 
N42  &  0 &  6.10 &  3.15 &  0.86 &  2.67 &  3.63 &  1.35 &  3.51 &  0.045 &  0.124 &  1.88 &  2.49 &  1.61 &  1.51 &       &       &       \\ 
N50  & -5 &  2.16 &  2.71 &  1.51 &  2.95 &  2.63 &  0.28 &  1.75 &  0.026 &  0.113 &  2.08 &  2.12 &  1.54 &  0.96 &       &       &       \\ 
     &    &       &       &       &       &       &{\it (0.78)} &   &        &        &       &       &       &       &       &       &       \\ 
\hline
\multicolumn{18}{l}{\underline{Background galaxies}} \\
N109 &  8 & -0.69 &  4.42 & -1.25 &  5.93 &  0.63 &  3.31 & -2.76 &  0.052 &  0.071 &  2.63 &  4.27 &  2.91 &  1.17 &  0.69 & -0.31 &  0.50 \\ 
N33  &  5 &  3.96 &  0.65 &  1.27 & -0.41 & -8.28 &  1.30 & -2.48 & -0.006 &  0.116 &  1.07 &  0.22 &  0.34 &  0.22 &  1.06 &       &       \\ 
N152 &  7 & -0.06 &  1.37 &  0.77 &  1.63 &  2.28 & -1.91 &  4.70 &  0.053 &  0.103 &  0.17 &  1.42 & -0.87 & -0.06 &  0.37 &       &       \\ 
N39  & -5 &  5.90 &  5.12 &  1.81 &  3.26 &  5.91 &  1.38 &  5.21 &  0.113 &  0.219 &  3.56 &  2.93 &  1.51 &  1.01 &  0.08 &       &       \\ 
B1   & -5 &  3.96 &  2.68 &  0.59 &  0.43 &  6.20 &  1.75 &  1.69 &  0.065 &  0.177 &  3.43 &  4.43 &  0.96 &  0.48 &  0.63 &       &       \\ 
B3   & -5 &  3.31 &  3.92 &  0.45 &  1.96 &  0.07 & -4.38 &  8.27 &  0.082 &  0.167 &  3.50 &  2.54 &  0.98 &  1.19 & -0.05 &       &       \\ 
\hline
\end{tabular}
\end{minipage}
\end{table*}

\subsection{Lick Indices}

In addition to the redshift information, the good spectroscopic material allowed
us to tackle in finer detail the study of spectral properties of member and background
galaxies in our sample. 
In particular, spectral resolution of both the {\sc Eso} and {\sc Casleo} 
observations closely matched the canonical prescription (i.e.\ $\sim 8$~\AA\ FWHM)
to consistently reproduce the Lick system \citep{worthey94}, thus allowing a wide set 
of narrow-band indices to be easily computed from the original data.

\begin{figure}
\centerline{
\includegraphics[width=0.91\hsize,clip=]{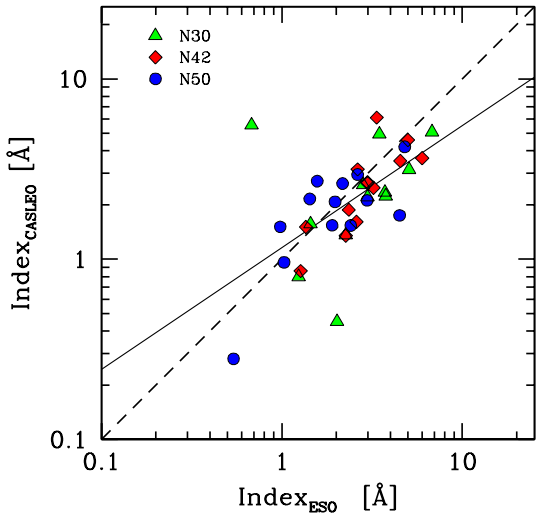}
}
\caption{
The set of Lick indices of Table~\ref{t8} for the three galaxies in
common between {\sc Eso} and {\sc Casleo} observations (namely, N30, N42, and N50)
Due to a poorer spectral resolution (10~\AA\ FWHM), the {\sc Casleo} spectra 
show slightly ``shallower'' spectral features (and correspondingly lower 
indices) compared to the {\sc Eso} spectra, better close to the Lick standard system. 
The dashed line in the plot is the one-to-one relation while the solid 
line is the least-squares fit of the points according to eq.~(\ref{eq:caseso}).
For better self-consistency, magnitude indices, like Mg$_1$ and Mg$_2$, 
have been converted to pseudo-equivalent widths, as explained in the
Footnote~\ref{foot12}.
}
\label{f14}
\end{figure}

As for the {\sc Eso} observations, our calculations have been carried out after slightly 
degrading the spectra to the Lick resolution by convolution with a Gaussian kernel. 
This transformation cannot be carried out with equivalent accuracy for the {\sc Casleo} data. 
Given their slightly poorer resolution (10~\AA~FWHM), 
in fact, these spectra tend to display shallower spectral features, and
therefore lower index strengths. The effect can be assessed by means of Fig.~\ref{f14}
for the three galaxies in common between {\sc Eso} and {\sc Casleo} spectra. The 
least-squares fit to the data indicates that the corresponding indices (in \AA\ pseudo-equivalent
width) relate as\footnote{To ease the comparison, in Fig.~\ref{f14} and
eq.~(\ref{eq:caseso}) both $Mg_1$ and $Mg_2$ indices
have been transformed from their magnitude scale to pseudo-equivalent width
recalling that, by definition, $I_{{\rm \AA}} = \Delta\,(1-10^{-0.4\,I_{\rm mag}})$,
being $\Delta$ the width of the feature window, according to the standard index
definition \citep{worthey94}. Therefore, we have that $\Delta = 65$~\AA\ for $Mg_1$ and
42.5~\AA\ for $Mg_2$.\label{foot12}}

\begin{equation}
\begin{array}{lll}
\log {\rm (I_{CAS})} = & \phantom{\pm}0.67\,\log {\rm (I_{ESO})} & +0.06\\
                    & \pm 0.15                        & \pm 0.07 \\
\end{array}
\label{eq:caseso}
\end{equation}
with $\rho = 0.60$ and $\sigma = 0.22$. 

\begin{figure}
\centerline{
\includegraphics[width=0.9\hsize,clip=]{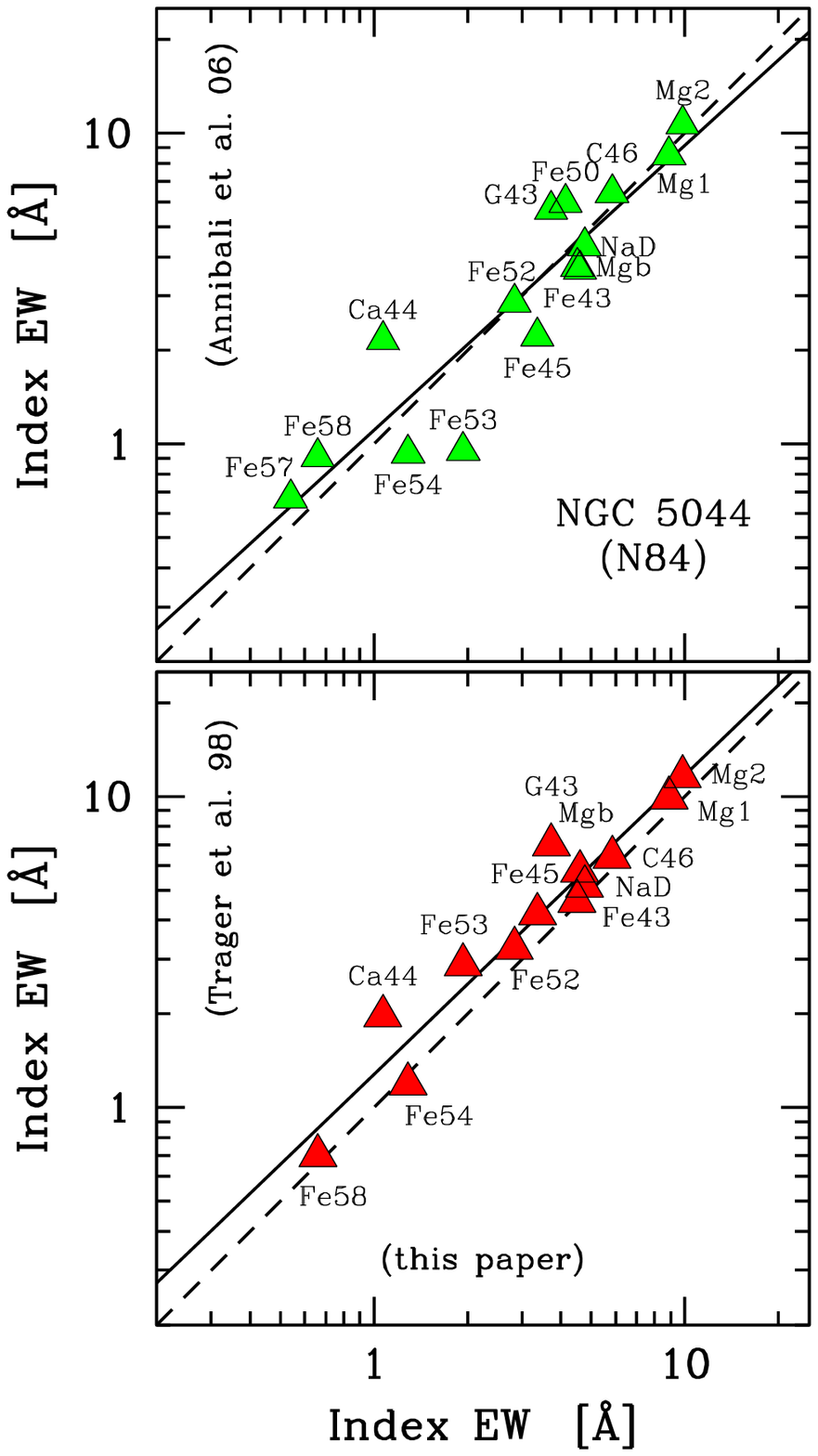}
}
\caption{
Lick-index comparison for NGC~5044. Our output from Table~\ref{t8} is matched with two 
calibrated sources in the literature for the same galaxy. In particular, the observations 
of \citet{annibali06} are reported in the upper panel, while the data of \citet{trager98} 
are displayed in the lower panel, as labelled on the vertical axis. $H\beta$ has been excluded 
in the plots being strongly affected by emission in this galaxy (see Fig.~\ref{f17}).
Like Fig.~\ref{f14}, for dimensional self-consistency, the Mg$_1$ and Mg$_2$ magnitude
indices have been converted here to pseudo-equivalent widths, as explained in the
Footnote~\ref{foot12}. Our data nicely correlate both with \citet{annibali06} ($\rho = 0.91$)
and with \citet{trager98} ($\rho = 0.96$). The scatter of our index residuals, namely
$\sigma[\log (I_{\rm our}/I_{\rm std})]$, with respect to the corresponding standard values
amounts to $\pm 0.15$~dex for \citet{annibali06} and $\pm 0.13$~dex for \citet{trager98}.
}
\label{f15}
\end{figure}

Although the lack of Lick primary calibrators in our sample prevented us
to fully standardize our index scale, nevertheless a direct comparison can be
done of our output for NGC~5044 itself with two reference sources in the 
literature, namely the work of \citet{trager98} and \citet{annibali06}.
As shown in Fig.~\ref{f15}, in both cases the index correlation is quite good
($[\rho, \sigma] = [0.96, 0.13]$ with Trager's data in the $\log$-$\log$
index domain and $[\rho, \sigma] = [0.91, 0.15]$ with the Annibali's ones)
assuring that the Lick standard system is correctly reproduced, on average, by our
observations. Table~\ref{t8} gives a general summary of our results.

%%%%%%%%%%%%%%%%%%%%

\begin{figure}
\centerline{
\includegraphics[width=\hsize,clip=]{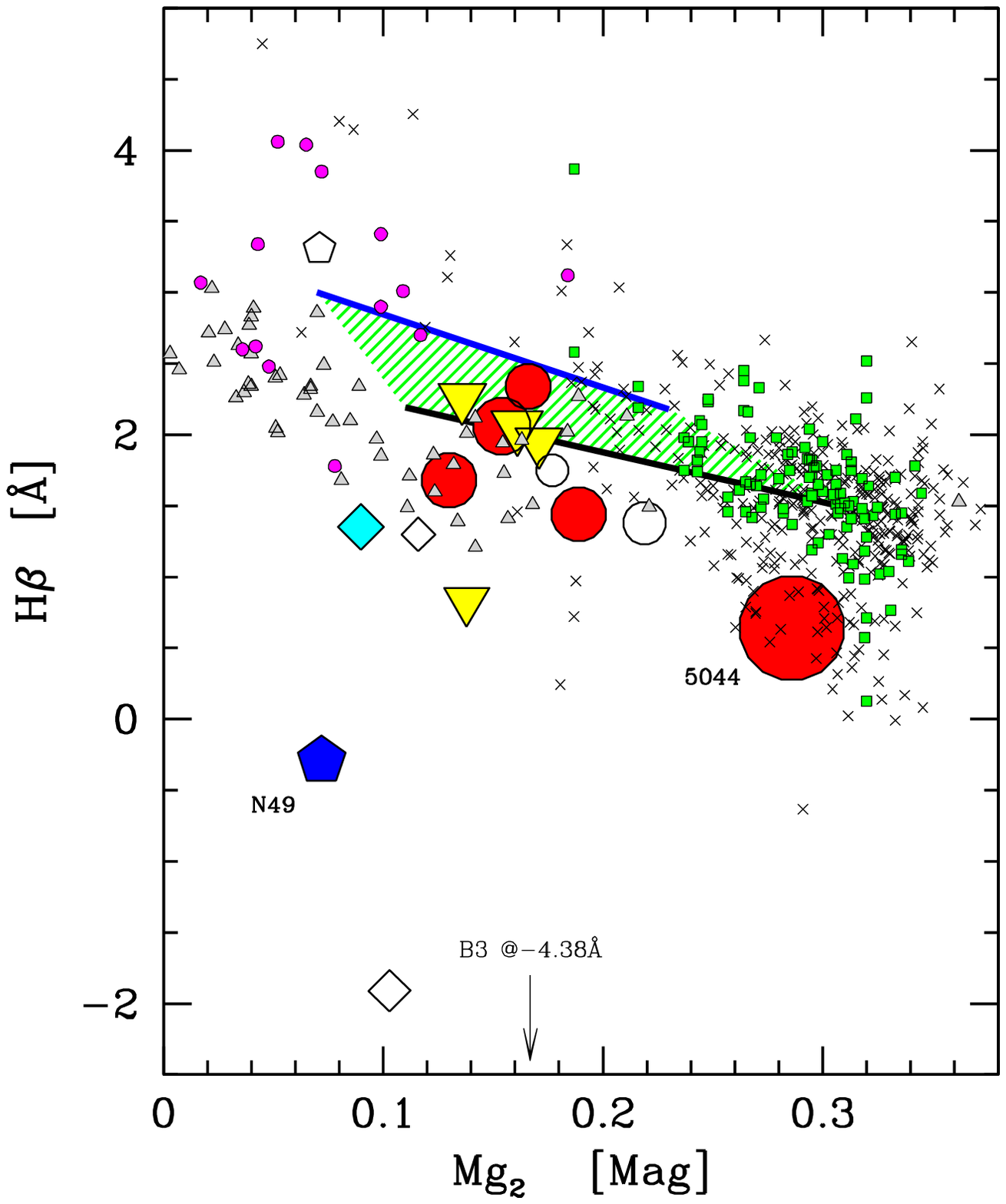}
}
\caption{
The Lick-gram of H$\beta$ vs.\ Mg$_2$ for the 16 {\sc Eso} galaxies (10 members +6 
background objects, the latter marked by open symbols) plus N29 from {\sc Casleo}. 
The small correction for H$\beta$ emission has been applied to the index of 
N17, N50 and N84, as indicated in Table~\ref{t8}. 
Marker size is proportional to $g$ luminosity, while morphological type
is traced by different symbols, as in Fig.~\ref{f6}.
The shaded region collects the theoretical SSP models of \citet{buzzoni92} and 
\citet{buzzoni94} for an age of 5 and 15 Gyr (upper and lower envelope, marked by 
thick solid lines), along the metallicity range $-1.3 \le [Fe/H] \le +0.25$ (in the sense of
increasing Mg$_2$ values).
The index distribution for the sample of 50 old M31 and Galactic globular cluster 
(small triangles) and 370 standard ellipticals (small crosses) from \citet{trager98} 
is also reported, for comparison, together with the group of 108 ellipticals with mild 
emission lines from \citet{rampazzo05} and \citet{annibali06} (heavy small squares) and 
14 young globular clusters of the Magellanic Cloud, according to \citet{freitas98} (small dots).
}
\label{f16}
\end{figure}

\subsubsection{H$\beta$ versus Magnesium}

An instructive view can be gained for the NGC~5044 group (and its surrounding 
background galaxies) in the Lick-index domain. The Magnesium Mg$_2$ index, together with 
the Balmer H$\beta$ strength are certainly among the most popular reference tracers for 
this kind of analysis for their better dependence on metallicity (Mg$_2$)
and age (H$\beta$), as extensively studied in the literature 
\citep[see, e.g.][for a discussion]{gorgas90, buzzoni95, thomas03, tantalo04}. 
The distribution of our sample is 
displayed in Fig.~\ref{f16}, comparing with the theoretical expectations for simple
stellar population models \citep{buzzoni92,buzzoni94} along an age range
between 5 and 15 Gyr, and with metallicity spanning the interval 
$-1.3 \le [Fe/H] \le +0.25$.
The sample of 50 old M31 and Galactic globular clusters, and 370 standard ellipticals from 
the work of \citet{trager98} is also superposed to the plot, together with a supplementary 
sample of 108 ellipticals with mild emission lines from \citet{rampazzo05} and \citet{annibali06}
for a differential comparison with the distribution of high-mass systems likely experiencing 
some moderate star-formation activity.
As a guideline for the distribution of young (metal-poor) stellar systems, we also 
added to the plot the sample of 14 globular clusters belonging to the Magellanic Cloud 
systems, according to \citet{freitas98}. 

As expected, the Mg$_2$-H$\beta$ diagnostic is very poor for late-type galaxies,
for which the H$\beta$ index is strongly affected by gas emission;
the location of the Im galaxy N49 in the plot is illustrative in this sense,
once considering its strong spectral emission, as in Fig.~\ref{f13}.
As far as the early-type galaxy component is concerned, however, 
one has to remark a pretty clean distribution of our dE+dS0 sample, fully intermediate
between Magellanic globular clusters and standard ellipticals, and pointing to a 
low (sub-solar) metallicity and an old age, consistent with the Hubble time.

\begin{figure}
\centerline{
\includegraphics[width=\hsize,clip=]{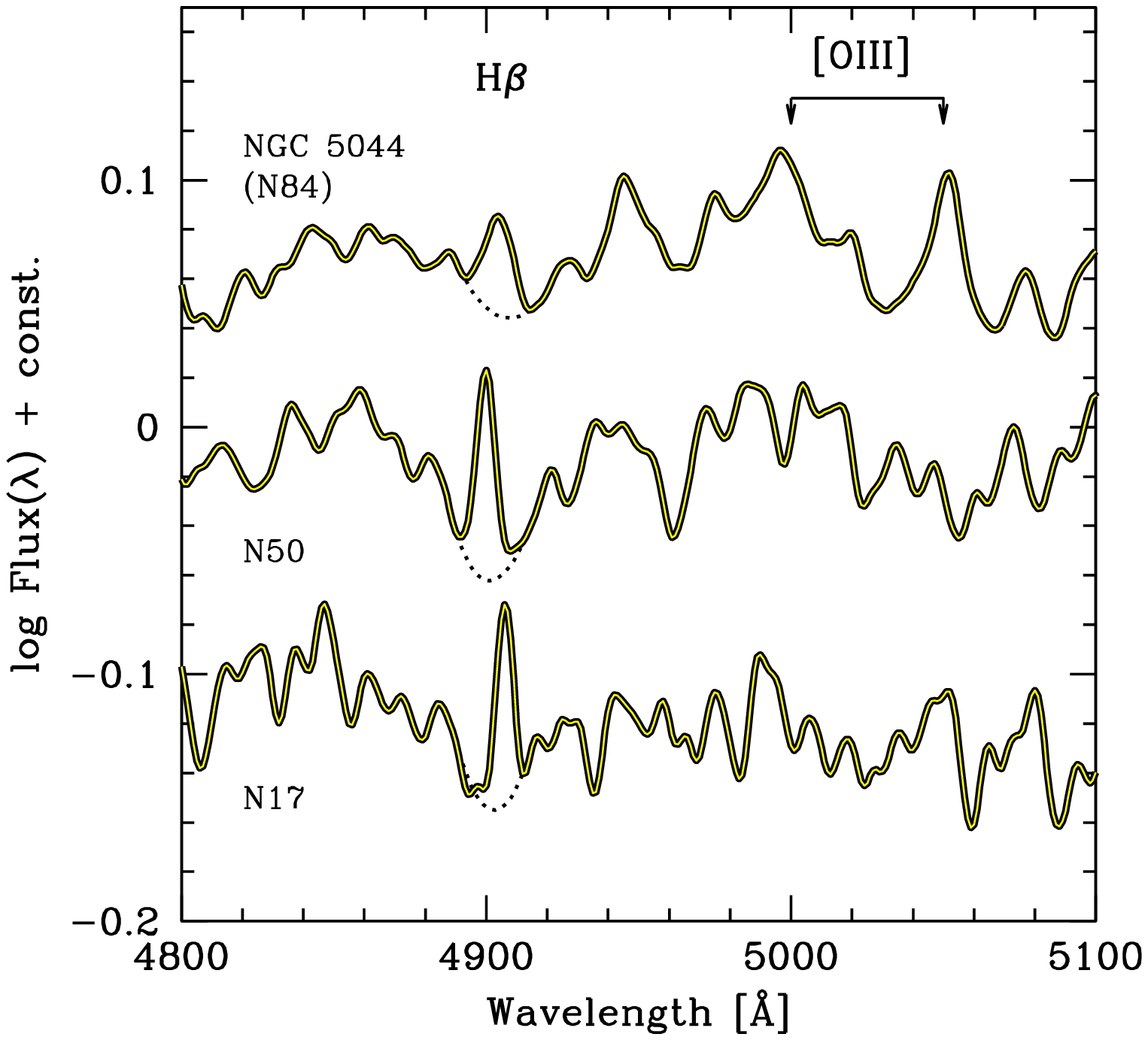}
}
\caption{
An H$\beta$ gas emission component, superposed to the stellar absorption
is clearly evident in the spectra of the three ellipticals NGC~5044 (alias N84), 
N50 and N17.
Note, for the latter cases, that no sizeable [O{\sc iii}] emission
is detected at the expected (restframe) wavelength of 4959 and 5007~\AA,
as labelled. The envisaged (conservative) correction to H$\beta$, as explained 
in the text, is indicated here by the dashed curves.
}
\label{f17}
\end{figure}
%%%%%%%%%%%%%%%%%%%%%%

In the latter respect, however, a word of caution still imposes in our conclusions
as even for these galaxies a somewhat peculiar H$\beta$ strength might be envisaged
to a closer analysis of the individual spectra (see Fig.~\ref{f17}).
NGC~5044 itself (alias N84 in our catalog classification) is certainly a case
in this regard, displaying an evident emission component, likely related to residual 
star formation and/or to a low-luminosity AGN 
\citep[see][and references therein]{gastaldello09,brough07}, superposed to the H$\beta$ absorption 
bulk.
The same pattern also appears in the spectra of galaxies N17 (dS0) and N50 (dE). 
For the latter object, in particular, a composite 
stellar population spanning a wide age range has been suggested by \citet{cb01} 
from a thorough study of the galaxy surface brightness distribution. Allover,
these data show that the presence of excited residual gas, triggered by 
fresh star formation, might be a somewhat pervasive condition 
marking the evolution of low-mass ellipticals even at present day.

A proper correction of the H$\beta$ index such as to single out the genuine
stellar absorption has widely proved, in the past, to be not a simple task.
A standard procedure has been devised by \citet{gonzalez93}, relying
on the parallel measurement of the [O{\sc iii}]$_{5007}$ strength, taken
as a ``proxy'' of the intrinsic H$\beta$ emission.\footnote{In case a significant
equivalent width E$_{{\rm [OIII]}}$ could be appreciated for the [O{\sc iii}]$_{5007}$ 
forbidden emission, then \citet{gonzalez93} suggested to enhance 
the observed value of H$\beta$ by $\Delta H\beta = 0.7 E_{{\rm [OIII]}}$.}
However, such a relationship has found controversial evidence in the literature
\citep[see][for a range of opinions]{carrasco95,trager00,serven10}, as Hydrogen and Oxygen 
coupling might be not so univocally constrained within the ISM gaseous phase depending
on the range of thermodynamical conditions \citep[e.g.][]{osterbrok74}.
Galaxy N50 itself, in Fig.~\ref{f17}, provides an outstanding example in this
sense, as its spiked H$\beta$ emission does not seem to be accompanied by any evident 
[O{\sc iii}]$_{5007}$ counterpart. Further support on this line is offered by 
\citet[][see Fig.~14 therein]{mendel09}.

As sketched in Fig.~\ref{f17}, only for the macroscopic case of elliptical galaxies 
N17, N50 (both in the ESO and {\sc Casleo} spectra), and N84 we eventually adopted a 
plain and very prudent correction procedure to the H$\beta$ index, by means of a 
spline fitting such as to remove {\it at least} the visible emission spike. The corrected 
galaxy points have been plotted in Fig.~\ref{f16}, while their resulting H$\beta$ 
index is reported in Table~\ref{t8} as an italics entry. In any case, it is 
clear that any hidden residual emission would reduce the H$\beta$ strength 
thus leading to overestimate the galaxy age, especially in case of younger star-forming 
ellipticals.

\subsubsection{Iron versus $\alpha$-elements}

Together with H$\beta$ and Mg$_2$, the sub-set of Fe5270 and Fe5335 indices, 
tracing the Fe{\sc i} features close to the strong Mg absorption at 5170~\AA, 
complete the bulk of popular indices extensively used in literature 
for galaxy diagnostic.
In particular, while Mg$_2$ is naturally sensitive to the abundance of the $\alpha$ 
elements, the Fe{\sc i} indices provide a complementary piece of information
to probe the overall metallicity of a stellar population \citep{buzzoni09}.

As sometimes attempted in the past literature \citep[e.g.][]{idiart95,gorgas97,zhu10}, 
a better display of the data can be done by averaging the two Fe features, like in 
a composite index
\begin{equation}
\langle Fe\rangle_2 = (Fe5270+Fe5335)/2,
\label{eq:fe2}
\end{equation}
which actually sums up the equivalent width of both features.

\begin{figure}
\centerline{
\includegraphics[width=\hsize,clip=]{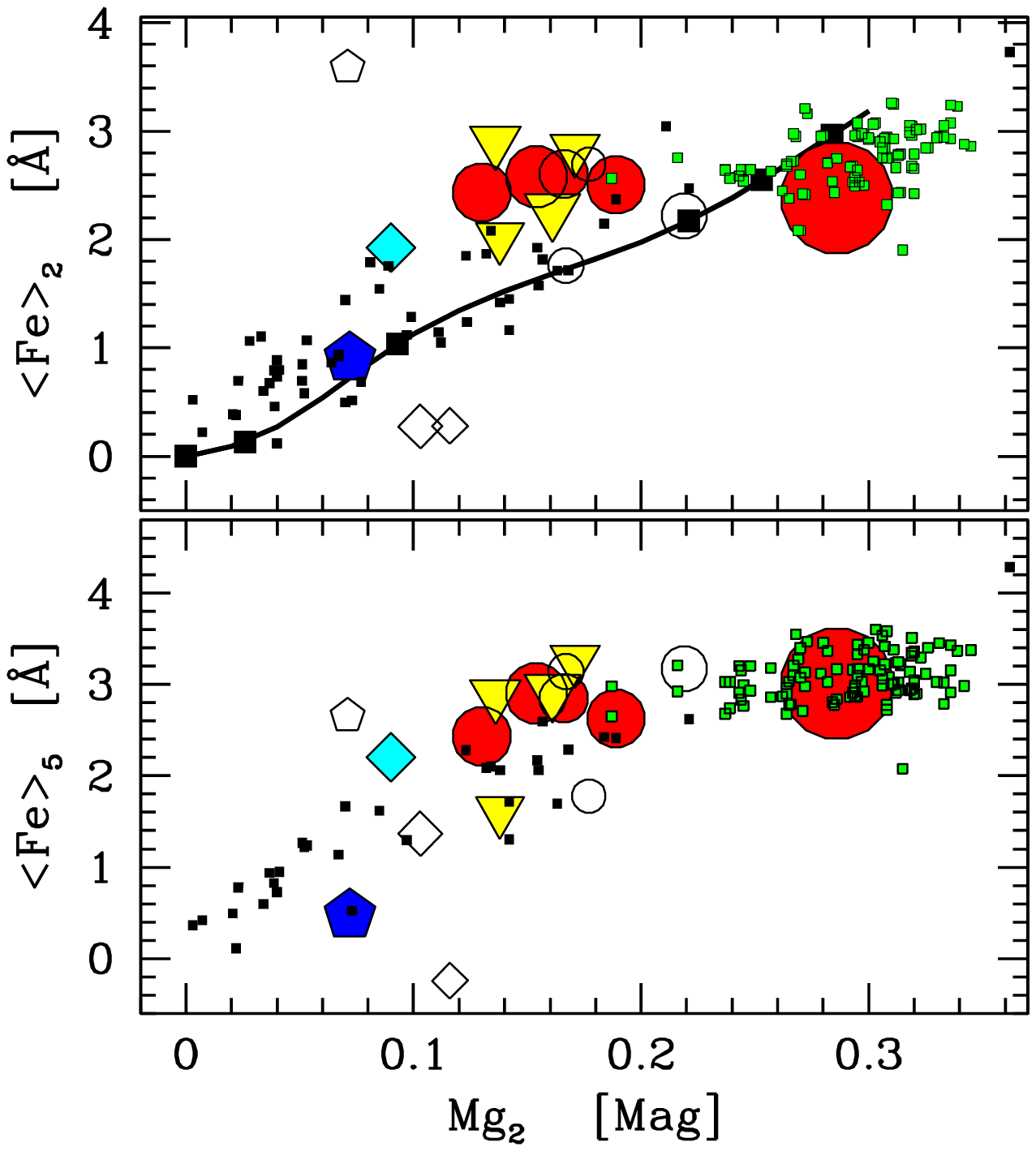}
}
\caption{
The Lick-gram of Iron meta-indices $\langle Fe\rangle_2$ {\it (upper panel)}
and $\langle Fe\rangle_5$ {\it (lower panel)}, as defined in the text
(eq.~\ref{eq:fe2}, and \ref{eq:fe6}) vs.\ Mg$_2$ for the 16 {\sc Eso} 
galaxies plus N29 from {\sc Casleo}. Symbols are as for Fig.~\ref{f16}.
The SSP theoretical locus for $t = 10$~Gyr, according to \citet{buzzoni92,buzzoni94},
is displayed (solid curve in the upper panel), with the metallicity values
marked up (solid squares) at [Fe/H] = $-\infty$, -2.3, -1.3, -0.25, $\odot$, +0.25,
in the sense of increasing Mg$_2$. 
The index distribution for the \citet{rampazzo05} and \citet{annibali06} homogeneous 
sample of ellipticals galaxies (small light squares) and the old M31 and Galactic 
globular clusters from \citet{trager98} (small dark squares) is also reported, for comparison.
}
\label{f18}
\end{figure}

An even cleaner result could also be secured for our dataset by taking advantage
of the full observation of the 5 ``Fe'' indices comprised in the Lick system within
the 4350-5406~\AA\ range (see, again, Table~\ref{t8}), excluding the
Fe5015 feature possibly affected in case of [O{\sc iii}]$_{5007}$ emission. A further composite 
index can therefore be built up as 
\begin{equation}
\begin{array}{rl}
\langle Fe\rangle_5 =& (Fe4383+Fe4531+Fe5270+\\
                     & +Fe5335+Fe5406)/5.
\end{array}     
\label{eq:fe6}
\end{equation}
Figure~\ref{f18} gives a summary of our results.
Again, for reader's better reference, our observed galaxy sample is matched 
in all the plots of the figure with the \citet{rampazzo05} and \citet{annibali06} standard 
ellipticals, as well as with the compilation of M31 and Galactic globulars according 
to \citet{trager98}.
The $\langle Fe\rangle _2$ index can also be easily computed for the theoretical models
of \citet{buzzoni94} and it is displayed in the upper plot of the figure as a
reference locus for 10 Gyr old SSPs to compare with the observed distribution.

As for the standard ellipticals, even our dwarf early-type galaxies display a 
flat $\langle Fe \rangle_2$ distribution vs.\ Mg$_2$.
This apparent decoupling between Iron and the other $\alpha$ elements (as traced
by Magnesium) is a well recognized feature \citep{gorgas90,worthey92,buzzoni94},
and probably the most direct evidence of the different enrichment channels
that provided metals to the galaxies in the past. According to stellar evolution 
theory, in fact, we know that $\alpha$-elements are important yields for high-mass 
stars ($M\gtrsim 8~M_\odot$) dying as Type II SNe; on the contrary, the Fe-Ni 
enrichment is more efficiently carried on by the Type Ia SNe, likely related to 
the binary-star environment. A steady trend of $\langle Fe \rangle_2$
vs.\ Mg might therefore be resilient of a constant abundance of Iron, that is
of a constant rate of Type Ia SNe \citep{buzzoni94}.

Apparently at odds with previous conclusions, however, the study of the $\langle Fe \rangle_5$ 
meta-index reports a more explicit correlation between Fe and Mg along the
entire mass range of standard and dwarf ellipticals.
To a finer detail, this puzzling behaviour is mostly induced by a trend in place
among the ``bluer'' Fe indices (namely Fe4383 and Fe4531). Contrary to 
the ``red'' indices (i.e.\ Fe5270, Fe5335 and Fe5406), when split into the different 
elemental contributions 
\citep[see, for instance, Table 2 in][]{trager98},
all the ``blue'' indices are actually blends including an important presence
of Ti and Mg, and this may eventually explain the apparent correlation 
between $\langle Fe \rangle_5$  and Mg$_2$.

As a final remark dealing with Fig.~\ref{f18}, one has to note the somewhat 
unexpected general correlation of galaxy indices along the different morhological types.
Although with larger individual uncertainties, in fact, also spirals and dwarf 
irregulars seem to obey in the plots the established relationship as for
ellipticals. This interesting behaviour is largely in consequence of the much
poorer (and similar) response of both Mg and Fe indices to SSP age, that simply
displaces their location in the plots along the SSP ``universal'' locus 
independently from the galaxy star-formation history.

\subsubsection{The $\alpha$-$\alpha$ element correlation}

\begin{figure}
\includegraphics[width=\hsize,clip=]{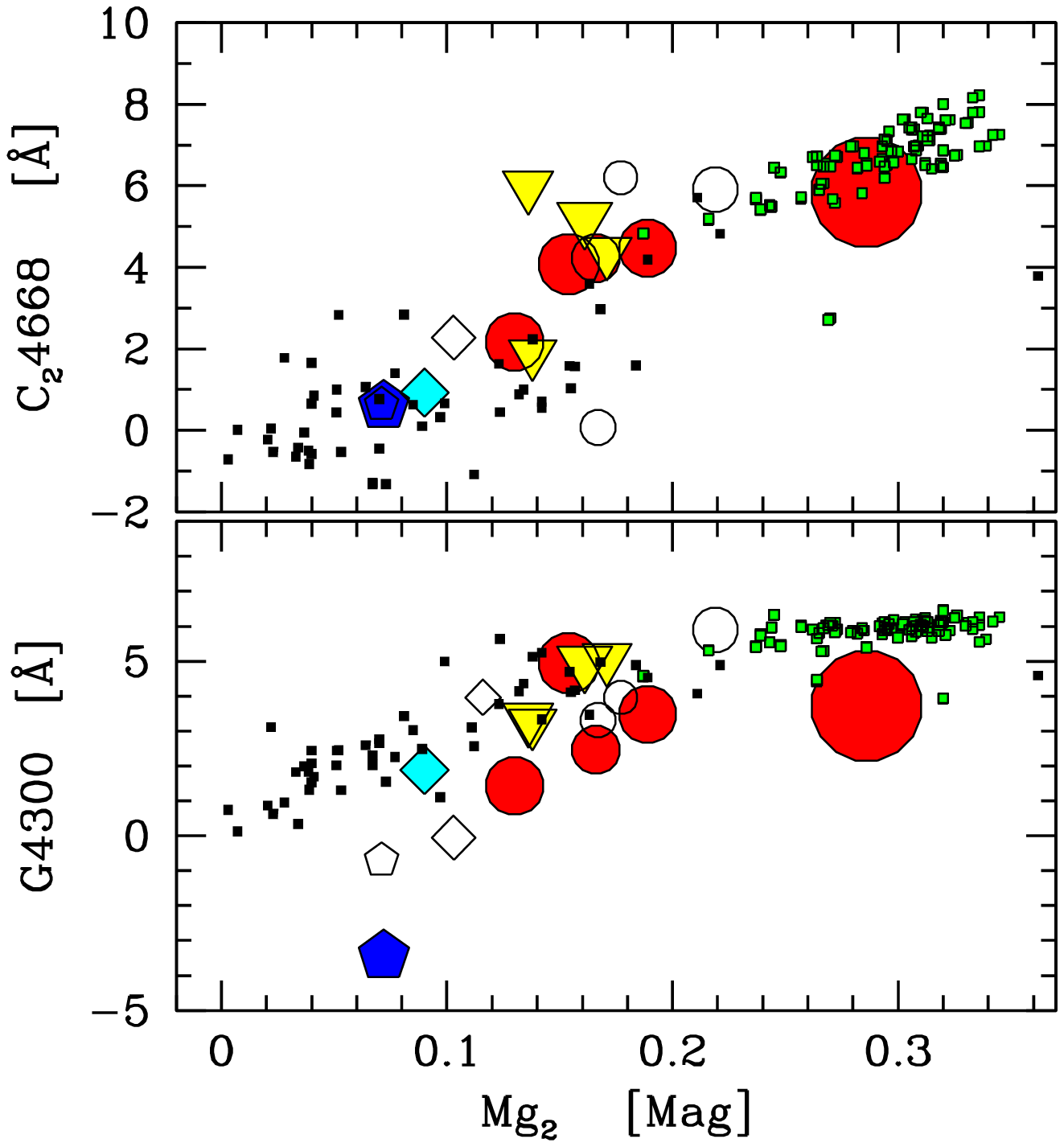}
\includegraphics[width=\hsize,clip=]{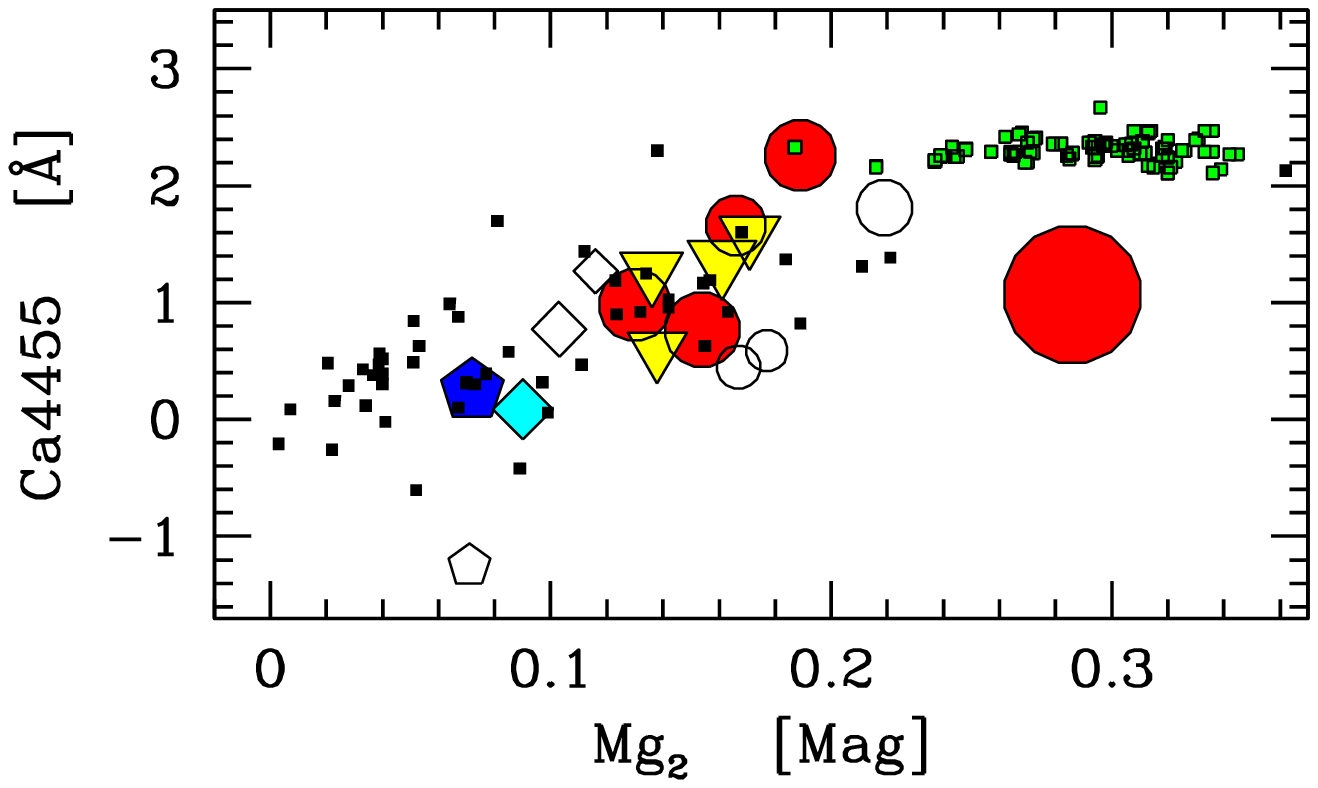}
\caption{
The Lick-gram for the two Carbon molecular indices
C$_2$ and CH (alias the G-band at 4300\AA), and for the Calcium index Ca4455
vs.\ Mg$_2$ for the 16 {\sc Eso} galaxies plus N29 from {\sc Casleo}. Symbols are as for 
Fig.~\ref{f16}. The \citet{rampazzo05} and \citet{annibali06} galaxy sample is also
overplotted, together with the M31 and Galactic globular clusters data of \citet{trager98}
(small squares). See text for a full discussion.
}
\label{f19}
\end{figure}

The natural correlation among $\alpha$ elements within our galaxy sample
can be verified by means of the two molecular features of Carbon, namely
CH (alias G4300) and C$_2$ at 4668~\AA\ (see the two upper panels of Fig.~\ref{f19}),
and the Calcium Ca4455 feature (as in the lower panel of the same figure). In both cases, a beautiful
trend is in place with the NGC~5044 dwarf-galaxy population linking standard 
ellipticals with globular clusters. A slightly more scattered plot may be noticed 
for the G4300 index, however, perhaps in some cases affected by the influence 
of the closeby H$\gamma$ emission.

Overall, one has also to report in the different panels of Fig.~\ref{f18} and \ref{f19}
a tendency for NGC~5044 itself to display slightly shallower absorption features
compared to the expected strength for standard ellipticals of similar
Mg$_2$ value. We are inclined to ascribe this effect to the larger velocity 
dispersion of this galaxy, which is by far the most massive one of the
group. The line broadening especially affects the shallowest features
lowering the corresponding index strength. This is especially true, for
instance, for the Ca4455 feature, as displayed in Fig.~\ref{f19}.
A hint in this sense also appears in Fig.~\ref{f15}, when comparing with
the standard indices of \citet{annibali06} and \citet{trager98}.

\section{Summary \& conclusions}

With this third paper of a series \citep[see also][]{cb01,cb05}, we conclude our 
long-term project on the study of the galaxy group surrounding the standard 
elliptical NGC~5044. As widely recognized in the recent literature 
\citep{faltenbacher05,sengupta06,brough06,mendel08,mendel09} this group
is an outstanding aggregate in the local Universe for its notable
population of low-mass (both dwarf and LSB) galaxies, which take part in a massive
($M_{\rm tot} \sim 1.6\,10^{13}~M_\odot$, \citealp{betoya06}) dynamically relaxed stucture, as traced 
by the diffuse X-ray emission of the group \citep{buote03,brough06}. 

Along several observing runs carried out at the {\sc Casleo} (San Juan, Argentina)
and {\sc Eso} (La Silla, Chile) telescopes we collected multicolour $B,g,V,r,i,z$ 
photometry for a significant fraction (79 objects) of the faintest galaxies projected 
on the NGC~5044 sky region assessing cluster membership on the basis of
apparent morphological properties (through accurate S\'ersic profile fitting)
and low-resolution ($R = 500-1000$) spectroscopy to estimate redshift for 25
objects in the field.

Both on the basis of the morphological and spectrophotometric properties of
the member galaxies, it is evident from our analysis a marked segregation
effect within the global population. As far as the galaxy surface-brightness profile
is concerned, a clear separation can be made in the S\'ersic parameter space
between early- and late-type systems, the first showing in nearly all cases
a ``concave'' (shape parameter $n \lesssim 1$) radial brightness profile.
Magellanic systems, on the contrary, stand out in the distribution for their
clean exponential profile ($n \simeq 1$, see Fig.~\ref{f3}), while dwarf
spheroidals confirm their sort of ``acephalous'' structure, where the
``convex'' profile ($n > 1$) lacks of any nuclear design, and the galaxy body
vanishes into a thiny external envelope of extremely faint mean surface brightness 
($\mu(g) \gtrsim 26$~mag arcsec$^{-2}$, see Fig.~\ref{f4}).
Overall, compared with a de Vaucouleurs $R^{1/4}$ profile, typical for 
standard ellipticals, the dwarf early-type galaxies in the group (dE, dS0, dSph) 
characterize for their ``broader'' (i.e.\ poorly-nucleated) surface-brightness 
distribution.

The intrinsically bluer colour and the larger colour spread, compared to
the other morphological sub-groups (see, in particular, Fig.~\ref{f5} and Table~\ref{t3}),
marks the population of Magellanic irregulars, pointing to 
a wide range of star-formation histories.
The case of the Sm galaxy N134 is a notable one in this regard, appearing to be
the youngest member of the N5044 group with an age well less than 1~Gyr (see Fig.~\ref{f8}),
probably as a result of its ongoing interaction with NGC~5054.
On the other hand, a drift toward bluer integrated colours is also an issue for 
dE's, a feature that may point to some moderate 
star-formation activity even among these nominally ``quiescent'' stellar systems
\citep[see, e.g.\ ][for similar results with the Fornax cluster dE's]{cellone96}.

Once rescaling for the group distance, the absolute magnitude distribution indicates that in the
NGC~5044 group we are dealing with a prevailing fraction of systems fainter than
$1.8\,10^9~L_\odot$ (see Sec.~4.2). In particular, dwarf ellipticals and irregulars, together, 
mark the bulk of the galaxy population around $M(g) \simeq -18.0\pm 1.5$, while
dSph's characterize the faint-end tail of the galaxy distribution (Fig.~\ref{f6} 
and \ref{f7}). The leading role
of disk galaxies may even better impose when considering that, apart from NGC~5044
that leads the group as an elliptical, the other three brightest members (heading then the
underlying dS0 line in the c-m diagram) are all spirals, namely NGC~5054 (type Sb), 
N68 (Sab) and N18 (Sab).

To some extent, this further emphasizes the quite special location of the NGC~5044 group,
within the cosmic aggregation scale, being strategically placed just ``midway'' between 
the high-density environment of galaxy clusters, and the low-density conditions 
of looser galaxy clumps like our Local Group. Recalling the recent discussion of
\citet{gavazzi10} about the ``nurture'' effects on galaxy evolution, it is interesting 
to remark that this ``duality'' of the NGC~5044 group reflects also in the prevailing 
galaxy mix within the aggregate, where the gas-depleted population of dwarf 
ellipticals (a typical sign of dense cosmic environments) coexists with an 
important population of dwarf (gas-rich) irregulars (the typical inhabitants of 
low-density regions in the Universe).

A tentative assessment of the mass budget sampled by the group member galaxies has been 
carried out in Sec.~4.2 on the basis of the observed colours and the morphology of 
each object. By relying on the \citet{buzzoni05} template models of Table~\ref{t4}, 
this combined piece of information allowed us to assign a representative M/L such 
as to convert absolute magnitudes into bright (baryonic) mass. Overall, the 63 
member galaxies in our sample complemented by other 50 likely members comprised in the
\citet{fs90} original catalog (which includes NGC~5044 and 5054 themselves),
and by 23 additional member galaxies with available photometry and morphological type from the
\citet{mendel08} survey (see Table~\ref{t5} and \ref{t6}) collect a total 
of $M_{\rm tot}^{\rm bright} = 2.3\,10^{12}$~M$_\odot$, with a mean representative value of 
$\langle \log M^*_{\rm gal}\rangle = 9.2 \pm 1.0$ for the member galaxies
(see Fig.~\ref{f9}). Roughly one fourth of $M_{\rm tot}^{\rm bright}$
is stored in NGC~5044 itself, while the three brightest members, alone, represent
nearly half the total bright mass of the group.

The derived total bright mass is about one seventh of the total dynamical mass of the group
as inferred from the X-ray emission map. This fraction might however be getting even smaller if 
one accounts for a larger estimate between $M_{\rm tot}^{\rm dyn}\sim 3.7\,10^{13}~M_\odot$
\citep{gastaldello07b} and $9.9\,10^{13}$~M$_\odot$, according to \citet{mendel08}, 
just on the basis of the galaxy velocity dispersion.

Like in the luminosity function, a clear morphological segregation
is also in place for galaxy masses with dE and dS0 systems which surmount the
Im and dSph component. The latter, in particular, constrains the
low-mass tail of the mass function providing the key connection between 
galaxies and globular clusters, around the $10^7$-$10^8$~M$_\odot$ mass range.
A more marked presence of later-type systems among the faintest 
galaxy population naturally complies with the downsizing mechanism \citep{gavazzi10}
making active star formation to better confine among low-mass aggregates, 
as shown in Fig.~\ref{f10}.

As discussed in Sec.~4, fresh stars are produced within the NGC~5044 group at a 
rate of roughly 23~M$_\odot$~yr$^{-1}$, implying a global birthrate $b \sim 0.15$.
This figure might however be a prudent lower limit, as it does not account
for the contribution of the two brightest members of the group and for the possible 
star-forming activity even among dE and dS0 galaxies. Some evidence in this 
sense stems, in fact, from the broad colour spread of these galaxies in 
Fig.~\ref{f8}, but even clearer signs emerge from the spectroscopic analysis 
of Sec.~5. In particular, Fig.~\ref{f17} shows that the general presence of 
H$\beta$ emission in the spectra of many dwarf ellipticals (and in NGC~5044 itself)
is resilient of a moderate but pervasive activity of fresh star formation
throughout in the group environment.

The good spectroscopic material, collected along the same photometric observing
runs added a number of useful details to characterize the evolutionary status
of the NGC~5044 group. These data allowed a univocal membership 
classification in most ambiguous cases discriminating between genuine dwarf
galaxies belonging to the group and bright background galaxies (see Table~\ref{t7});
the case of objects B3 and N39 is illustrative in this sense, as both 
ellipticals eventually resulted to be part of a background structure at 
$z \sim 0.095$. In this regard, at least three relevant galaxy aggregates 
appear beyond our group, the farthest one being a confirmed galaxy cluster at 
$z \sim 0.28$ \citep{gastaldello07}.

The resolution and wavelength coverage for {\sc Eso} and {\sc Casleo} spectra also allowed us
to self-consistently derive a series of Lick narrow-band indices, as summarized in
Table~\ref{t8}. In particular, eight relevant features of Iron between
4300 and 5800~\AA\  have been sampled, while $\alpha$ elements can be traced 
by means of the 4455~\AA\ Ca feature and the striking Mg feature
at 5170~\AA. In addition, two molecular bands of Carbon (CH at 4300~\AA, 
and C$_2$ at 4668~\AA) complete our diagnostic tools.
As far as early-type galaxies are concerned, spectroscopic analysis clearly 
indicate for them a sub-solar metallicity ($-1.0 \lesssim [Fe/H] \lesssim -0.5$,
see especially Fig.~\ref{f18}), and in general an old age, consistent with the
Hubble time. In the latter respect, however, one has to be aware of the possible hidden
bias (toward higher values) in age diagnostic via H$\beta$ absorption strength 
(Fig.~\ref{f17}) due to the effect of gas emission.

As for the established relationship for high-mass ellipticals \citep{worthey92}, 
also the dwarf population of NGC~5044 shows a decoupled trend between Iron and 
$\alpha$ elements (as in the uppel panel of Fig.~\ref{f18}), consistent with the mild 
trend between $[\alpha/Fe]$ and stellar mass for $M > 10^{8.5}~M_\odot$ galaxies shown by
\citet{mendel09}.
This feature supports and further extends also to the low-mass framework the canonical 
picture for galaxy chemical evolution as a result of the two distinct enrichment
channels dealing with Type I and Type II SNe \citep[see, e.g.][]{matteucci86}.

\section*{Acknowledgments}
We warmly thank {\sc Eso} and {\sc Casleo} staffs for their skillful assitance. 
The anonymous referee is also acknowledged for several timely comments to the
original draft. 
Funding from ANPCT (Argentina), PICT 03-00339 (1998) is also acknowledged.
SAC would like to thank CONICET and UNLP for funding through personal grants, 
and the Astronomical Observatories of Brera-Merate and Bologna, Italy, for their 
hospitality. The Universidad Nacional de La Plata is also acknowledged by one of 
us (AB) for friendly and generous support along the many visits to Argentina 
in the framework of this project.

%\clearpage
\appendix
\section{Relevant properties of the NGC~5044 group galaxies}

We summarize in Tables~\ref{a2} and \ref{a3} the main distinctive properties
of the observed galaxy population in the NGC~5044 group.
Column caption, for both tables is as follows:
\begin{description}\itemsep4pt
\item[col.~1] - Galaxy ID (see, in this regard, Footnote~\ref{footn});
\item[col.~2] - Membership code, $m_c$: 1~=~definite; 2~=~likely; 3~=~possible; 4~=~non member.
Membership is assigned according to: {\it i)} $cz < 3800$~km\,s$^{-1}$ if redshift is 
available from either CB05 and/or \citet{mendel08} and/or other),{\it ii)} morphology 
(if no redshift available).
\item[col.~3] - Morphological classification according to CB05. See \citet{cellone07} for
the meaning of ``mixed'' morphological classes;
\item[col.~4] - Morphological classification according to FS90;
\item[col.~5] - De Vaucouleurs' morphological type, $T$;
\item[cols.~6 and 7] - Isophotal radius ($\rho_{27}$) at $\mu = 27$~mag\,arcsec$^{-2}$, and 
effective radius ($\rho_\mathrm{e}$), which encircles 50\% of the luminosity within $\rho_{27}$.
The $g$-band imagery is taken as a reference for the {\sc Eso} sample, and 
$V$-band for the {\sc Casleo} galaxies;
\item[cols.~8,9, and 10] - S\'ersic fitting parameters from $g$-band ({\sc Eso} sample)
and $V$-band imagery ({\sc Casleo} sample). These are central brightness ($\mu_0$)
in mag arcsec$^{-2}$, scale radius ($\rho_0$) in arcsec, and ``shape index'' $n$;
\item[col.~11] - Total apparent magnitude encircled within the $\mu = 27$mag\,arcsec$^{-2}$ 
isophote in the $g$ ({\sc Eso}) and $V$ ({\sc Casleo}) bands;
\item[col.~12] - Mean surface brightness within $\rho_{27}$ isophotal radius in the $g$ 
({\sc Eso}) and $V$ ({\sc Casleo}) bands;
\item[col.~13] - Mean surface brightness within one effective radius $\rho_\mathrm{e}$ in the $g$ 
({\sc Eso}) and $V$ ({\sc Casleo}) bands;
\item[col.~14 to 19] (Table~\ref{a2}) - Gunn $griz$ colours within the $g$-band values of 
$\rho_{27}$ and $\rho_\mathrm{e}$ for the {\sc Eso} galaxies;
\item[col.~14 to 15] (Table~\ref{a3}) - Johnson $BV$ colours within the $V$-band values of 
$\rho_{27}$ and $\rho_\mathrm{e}$ for the {\sc Casleo} galaxies. 
\end{description}

\clearpage 

\begin{table*}
\begin{minipage}{200mm}
\caption{Morphological and photometric properties of the {\sc Eso} galaxy sample}
\label{a2}
%\vspace{30mm}
\end{minipage}
\end{table*}

\begin{figure*}
\includegraphics[width=0.62\hsize,clip=]{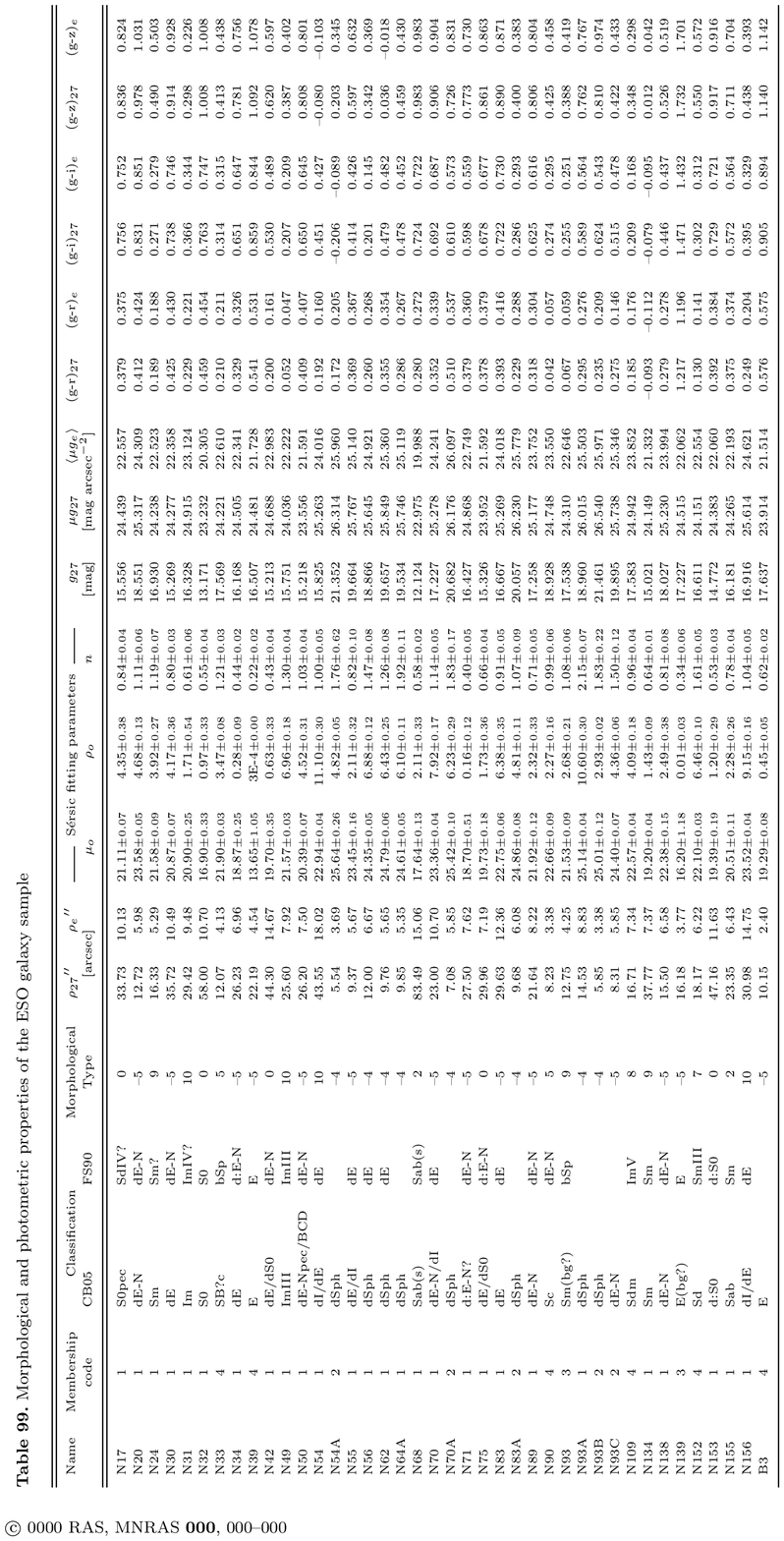}
\end{figure*}

\begin{table*}
\begin{minipage}{200mm}
\caption{Morphological and photometric properties of the {\sc Casleo} galaxy sample}
\label{a3}
%\vspace{30mm}
\end{minipage}
\end{table*}

\begin{figure*}
\includegraphics[width=\hsize,clip=]{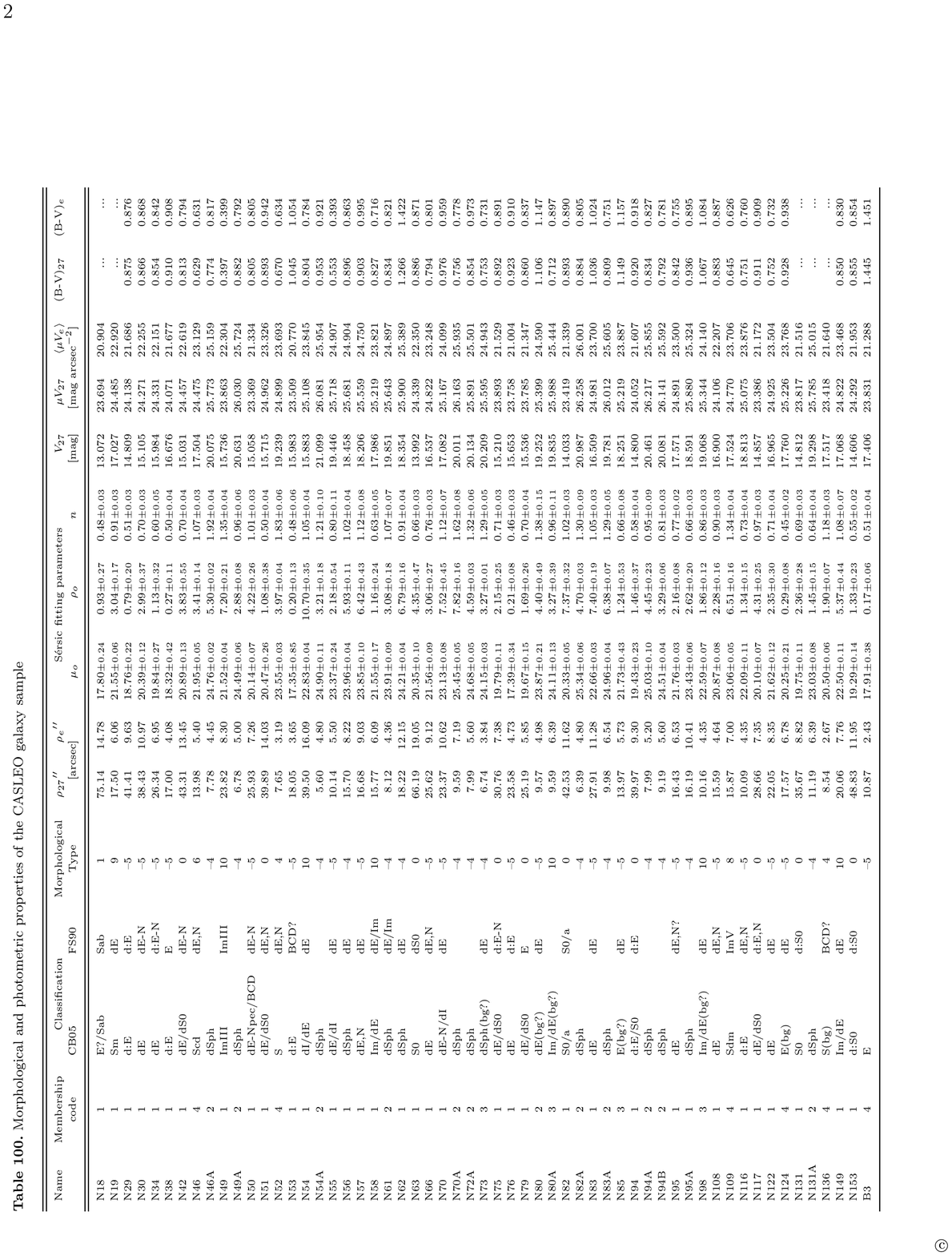}
\end{figure*}

\clearpage

\begin{table}
\caption{Coordinates of the 6 new dSph galaxies discovered in the {\sc Casleo} fields}
\label{a1}
\begin{tabular}{lcc}
\hline\hline
Name & $\alpha_\mathrm{J2000}$ & $\delta_\mathrm{J2000}$ \\
\hline
  N46A   &   $13^{\rm h}\,14^{\rm m}\,19.1^{\rm s}$ &  $-16\degr\,17'\,32''$ \\
  N72A   &   $13^{\rm h}\,14^{\rm m}\,51.8^{\rm s}$ &  $-15\degr\,56'\,28''$ \\
  N80A   &   $13^{\rm h}\,15^{\rm m}\,11.8^{\rm s}$ &  $-16\degr\,59'\,48''$ \\
  N82A   &   $13^{\rm h}\,15^{\rm m}\,23.4^{\rm s}$ &  $-16\degr\,27'\,18''$ \\
  N94A   &   $13^{\rm h}\,15^{\rm m}\,34.2^{\rm s}$ &  $-16\degr\,30'\,36''$ \\
  N94B   &   $13^{\rm h}\,15^{\rm m}\,39.7^{\rm s}$ &  $-16\degr\,29'\,56''$ \\
\hline
\end{tabular}
\end{table}

\begin{table}
\caption{Coordinates of background ``bonus'' galaxies.}
\label{a4}
\begin{tabular}{lcc}
\hline\hline
Name & $\alpha_\mathrm{J2000}$ & $\delta_\mathrm{J2000}$ \\
\hline
%   ID    x      y          RA        Dec    g_bes r_bes i_bes z_bes g_isc r_isc i_isc z_isc g_ap  r_ap  i_ap  z_ap  g_is   r_is  i_is  z_is    -------------------------------------errori associati-------------------------------------     Fgiso   Fr_is    Fi_is  Fz_is   Area   --errori---flussi--   x2       y2      xy     a  theta    mug   mur    mui   muz    flag   fwhm  elon  class
B1 & $13^{\rm h}\, 14^{\rm m}\, 19.0^{\rm s}$ & $-16\degr\, 10'\, 38''$ \\
B2 & $13^{\rm h}\, 15^{\rm m}\, 02.1^{\rm s}$ & $-16\degr\, 22'\, 17''$ \\
B3 & $13^{\rm h}\, 16^{\rm m}\, 07.4^{\rm s}$ & $-17\degr\, 00'\, 08''$ \\
B4 & $13^{\rm h}\, 17^{\rm m}\, 41.9^{\rm s}$ & $-16\degr\, 10'\, 07''$ \\
B5 & $13^{\rm h}\, 14^{\rm m}\, 00.0^{\rm s}$ & $-15\degr\, 56'\, 35''$ \\
B6 & $13^{\rm h}\, 17^{\rm m}\, 42.7^{\rm s}$ & $-16\degr\, 32'\, 48''$ \\
\hline
\end{tabular}
\end{table}

\bsp
\label{lastpage}
\end{document}